\documentclass[%
 reprint,
 superscriptaddress,
 amsmath,amssymb,
 aps,
prb,
]{revtex4-2}

\usepackage{graphicx}
\usepackage{dcolumn}
\usepackage{bm}
\usepackage{chemformula} 
\usepackage{times} 
\usepackage{tabularx}
\usepackage{physics}
\usepackage{hyperref}
\usepackage{mathtools} 
\usepackage{xcolor}
\usepackage{amsmath} 

\usepackage{chngcntr}

\begin{document}

\preprint{APS/123-QED}

\title {Magnetic excitations and interactions in the Weyl ferrimagnet NdAlSi} 

\author{Chris J. Lygouras}
\affiliation{Institute for Quantum Matter and William H. Miller III Department of Physics and Astronomy, Johns Hopkins University, Baltimore, Maryland 21218, USA}

\author{Hung-Yu Yang}
\affiliation{Departments of Physics, Boston College, Chestnut Hill, 02467, MA, USA}

\author{Xiaohan Yao}
\affiliation{Departments of Physics, Boston College, Chestnut Hill, 02467, MA, USA}

\author{Jonathan Gaudet}
\affiliation{Institute for Quantum Matter and William H. Miller III Department of Physics and Astronomy, Johns Hopkins University, Baltimore, Maryland 21218, USA}
\affiliation{NIST Center for Neutron Research, National Institute of Standards and Technology, Gaithersburg, Maryland 20899, USA}
\affiliation{Department of Materials Science and Engineering, University of Maryland, College Park, Maryland 20742-2115, USA} 

\author{Yiqing Hao}
\affiliation{Neutron Scattering Division, Oak Ridge National Laboratory, Oak Ridge, Tennessee 37831, USA}

\author{Huibo Cao}
\affiliation{Neutron Scattering Division, Oak Ridge National Laboratory, Oak Ridge, Tennessee 37831, USA}

\author{Jose A. Rodriguez-Rivera}
\affiliation{NIST Center for Neutron Research, National Institute of Standards and Technology, Gaithersburg, Maryland 20899, USA}
\affiliation{Department of Materials Science and Engineering, University of Maryland, College Park, Maryland 20742-2115, USA} 

\author{Andrey Podlesnyak}
\affiliation{Neutron Scattering Division, Oak Ridge National Laboratory, Oak Ridge, Tennessee 37831, USA}

\author{Stefan Bl{\"u}gel} 
\affiliation{Peter Grünberg Institut and Institute for Advanced Simulation, Forschungszentrum Jülich \& JARA, D-52425 Jülich, Germany}

\author{Predrag Nikoli{\'{c}}}
\affiliation{Institute for Quantum Matter and William H. Miller III Department of Physics and Astronomy, Johns Hopkins University, Baltimore, Maryland 21218, USA}
\affiliation{Department of Physics and Astronomy, George Mason University, Fairfax, VA 22030, USA}

\author{Fazel Tafti}
\affiliation{Departments of Physics, Boston College, Chestnut Hill, 02467, MA, USA}

\author{Collin L. Broholm}
\affiliation{Institute for Quantum Matter and William H. Miller III Department of Physics and Astronomy, Johns Hopkins University, Baltimore, Maryland 21218, USA}
\affiliation{NIST Center for Neutron Research, National Institute of Standards and Technology, Gaithersburg, Maryland 20899, USA}
\affiliation{Department of Materials Science and Engineering, Johns Hopkins University, Baltimore, Maryland 21218, USA}

\begin{abstract} 
Weyl fermions can arise from time-reversal symmetry-breaking magnetism, but their impact on magnetic order is a source of ongoing research. Using high-precision neutron diffraction and spectroscopy, we present a comprehensive exploration of the magnetic structure and excitation spectrum of Weyl semimetal and helical magnet NdAlSi. We use Luttinger-Tisza, classical mean-field, and random-phase approximation techniques to model the dispersive crystal field excitons. We find extended-ranged and sign-changing interactions, suggesting a coupling between conduction electrons and the local moments. We demonstrate that low-symmetry anisotropic Dzyaloshinskii-Moriya interactions, in contrast with higher-symmetry interactions enabled by Weyl fermions, play an important role in stabilizing the complex spin spiral ground state of NdAlSi. Our work provides a first detailed view of microscopic interactions in a Weyl magnet, and constrains the role of Weyl electrons and their chirality on the spiral magnetism. 
\end{abstract} 

\date{\today}
\maketitle 

\section{Introduction} 
The notion of topology is imperative for a comprehensive understanding of the physics of many condensed matter systems. Unique phases of matter, like the topological semimetal, arise due to symmetry-protected degeneracies. Of these phases, Weyl semimetals are unique in that they harbor Weyl fermions as quasiparticles, subject to relativistic linear energy-momentum relation and a topologically-protected chirality. Weyl nodes, Weyl fermions at zero energy, are sources and sinks of Berry curvature in momentum space, acting as charged singularities in the form of magnetic monopoles \cite{Weng2015, Armitage2018}. Integrating the Berry curvature on a surface in the vicinity of the Weyl node yields a nontrivial Chern number, a topological charge characterising the chirality of the Weyl fermion, which can only change by annihilation with a Weyl node of opposite chirality. The Weyl semimetal arises by breaking time-reversal symmetry or inversion symmetry. In noncentrosymmetric Weyl materials, time-reversal and crystalline symmetry dictates the arrangement and number of Weyl nodes in the Brillouin zone to be at least four \cite{Halasz2012}. The presence of these singular Weyl nodes in the bulk has a drastic impact on electronic properties, including Fermi surface arcs \cite{Armitage2018, Xu2015, Xu2017}; the chiral anomaly manifesting in negative longitudinal magnetoresistance due to the nonconservation of chiral left- and right- moving fermions \cite{Nielsen1983, Xiong2015}; large linear transverse magnetoresistance due to linear band dispersion \cite{Abrikosov2000, Wang2011, Shekhar2015, Chamorro2019}; anomalous Hall effect generated from enhanced Berry curvature \cite{Nayak2016, Chen2020, Nagosa2010, Smejkal2022}; and singular magnetoresistance from nodal Fermi surfaces \cite{Suzuki2019}. 

While the classification scheme for non-interacting topological insulators and semimetals is well-established \cite{Fu2007, Young2012, Yang2014}, the role of interactions and strong correlations in these systems is the focus of ongoing research. In particular, how might these topological fermions influence and imprint upon collective phenomena? In recent years, there have been several advances in this area, including studies of: the Kohn anomaly due to Weyl electron-phonon coupling \cite{Drucker2023}; the impact of topological fermions on the superconducting gap structure and anisotropy \cite{Guguchia2017, YiLi2018, Lygouras2023}; the amplification of Dirac fermion physics from formation of charge density waves \cite{Lei2021}; Weyl-Kondo heavy fermion physics \cite{Dzsaber2021, Fuhrman2021}; and coupling between local moments and itinerant Weyl electrons \cite{Chang2018, Borisenko2019, Gaudet2021, Gaudet2023, ZhangNan2023, Yao2023}. 

\begin{figure}
    \centering
    \includegraphics[width=\linewidth]{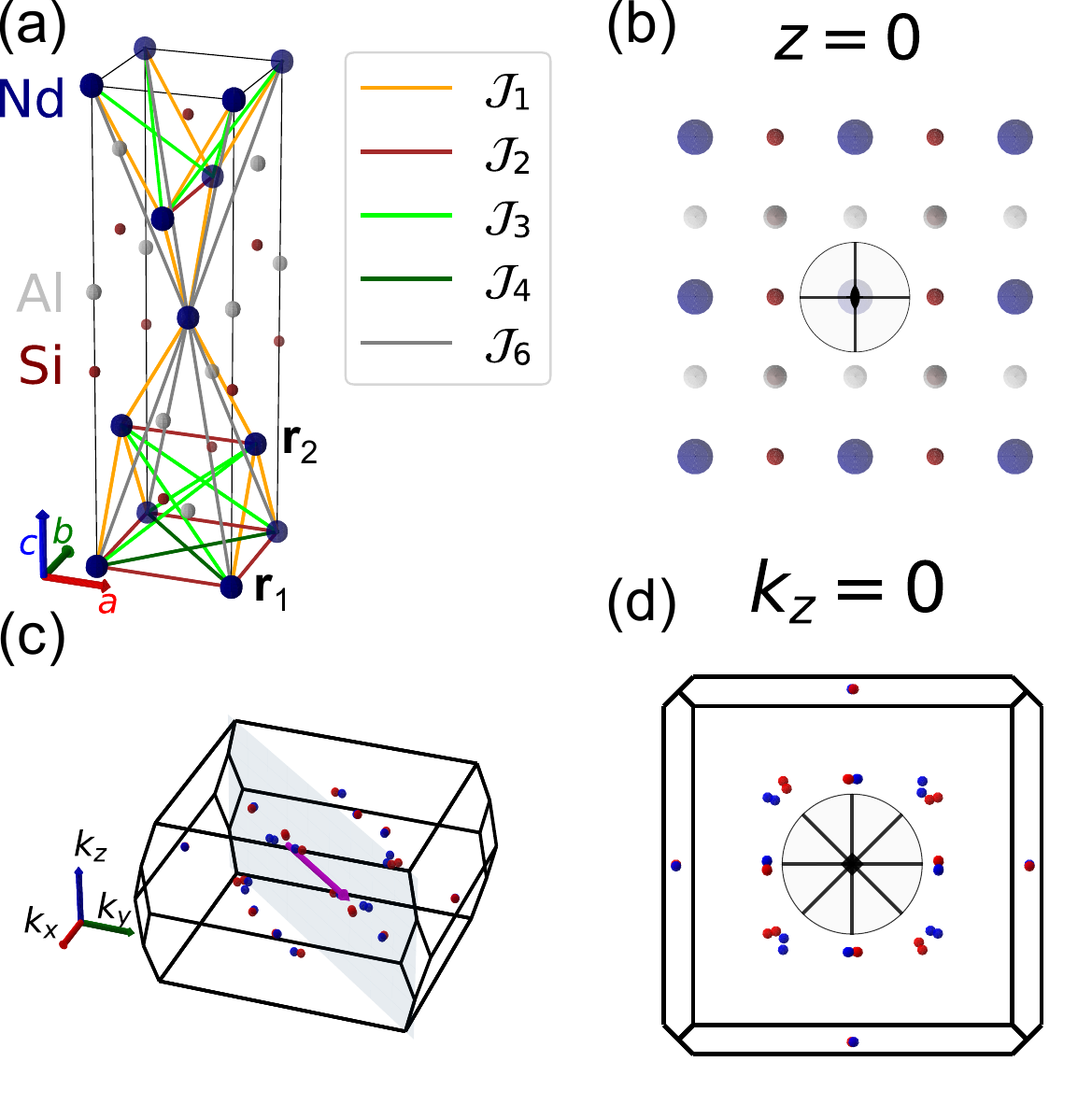}
    \caption{(a) Crystal structure highlighting the different elements, sites $\vb{r}_1,\vb{r}_2$, and bonds; and (b) atoms near the $z=0$ plane, highlighting the orthorhombic $C_{2v}$ symmetric bonds. (c) Brillouin zone and the $(hhl)$ scattering plane to which our neutron experiments are most sensitive; along with (d) the $k_z=0$ plane, highlighting the tetragonal $C_{4v}$ symmetry of Weyl points.}
    \label{fig:CrystalStructure}
\end{figure}

A prototypical family of inversion symmetry-breaking Weyl semimetal candidates is the $R$Al$X$ system, where $R$ is a rare-earth ion and $X$ is a tetral (Group 14) element, namely Si or Ge. This family consists of nominally non-centrosymmetric achiral crystals in space group 109 ($I4_{1}md$) with interpenetrating body-centered tetragonal unit cells related by nonsymmorphic symmetries with a translation $\vb{t} = (0, 1/2, 1/4)$. The Weyl fermions arise from intersecting valence and conduction bands and their number and distribution of the nodes depends on the mirror and rotation operators of the point group $C_{4v}$. LaAlGe was one of the first Lorenz-violating Type-II Weyl semimetal candidates as shown by angle-resolved photoemission spectroscopy (ARPES) and density functional theory (DFT) studies \cite{Chang2018, Xu2017}, and has recently been shown to display superconductivity at pressures beyond 65 GPa \cite{Cao2022}. The inversion-symmetry breaking is vital for the existence of Weyl fermions in this system. In principle, random alloying of Al and Si/Ge would restore the mirror plane $\sigma_z$ and restore centrosymmetry, bringing the crystal system to the space group 141 ($I4_{1}/amd$) which does not accommodate Weyl fermions \cite{Bouaziz2024}. Some X-ray diffraction studies have suggested the existence of one space group over the other, or their coexistence \cite{Pukas2004, Wang2021}; however, X-rays are not as sensitive to the noncentrosymmetry as neutron diffraction and second harmonic generation \cite{Yang2021, Gaudet2021}. Additionally, this system can accommodate magnetic rare-earth ions in place of La, from Ce through Gd \cite{Chang2018, Yang2020, Yang2021, Gaudet2021, Laha2024}. Fig.\ref{fig:CrystalStructure}(a) shows the crystal structure and various exchange interactions between magnetic ions. The introduction of magnetism into these Weyl semimetal candidates has opened doors to explore various interesting phenomena related to how differing magnetocrystalline anisotropy, isotropic and relativistic exchange interactions, total angular momentum quantum number, and more, might allow for observable effects of their interactions with Weyl fermions \cite{Yang2020, Gaudet2021, Wang2022, ZhangNan2023, Lou2023, Zhang2023, Yao2023}. 

The magnetic analog NdAlSi has proved an interesting material due to the potential connection between its magnetic ordering wavevector $\vb{k}_\mathrm{mag} \approx (0.67, 0.67, 0)$ and the coincident momentum-space Weyl-node separation $\Delta \mathbf{Q}_{\mathrm{Weyl}} \sim \vb{k}_\mathrm{mag}$ \cite{Gaudet2021}. Fig.\ref{fig:CrystalStructure}(c,d) show the distribution of Weyl points in the Brillouin zone (blue and red) and the magnetic wavevector (purple) which connects Fermi surface pockets that contain these Weyl points. The Kondo coupling between local moments and itinerant Weyl electrons can be rewritten in an effective theory as interactions between localized moments mediated by the Weyl fermions, in the form of a Ruderman-Kittel-Kasuya-Yosida (RKKY) interaction \cite{Ruderman1954, Nikolic2021, Chang2015}. Additionally, previous neutron diffraction data have revealed an incommensurate transition at $T_\mathrm{inc}\approx 7.2$ K to a long-wavelength modulated state, followed by a commensurate ordering at $T_\mathrm{com}\approx 3.5$ K to a nearly-Ising ferrimagnetic up-down-down ($udd$) spin structure with small canting angles into the basal plane \cite{Gaudet2021}. Weyl fermions, being chiral quasiparticles, may leave imprints of their chirality on the exchange interactions, in particular the Dzyaloshinskii-Moriya (DM) interaction, which can promote such helical spin structures. This presents a unique opportunity to explore how Weyl fermions may influence collective magnetic order. In recent works, it was shown that the ferrimagnetism in NdAlSi influences the electronic topology by renormalizing the band structure and the positions of the Weyl fermions \cite{Li2023}. Furthermore, the Weyl electrons and their nodal Fermi surfaces have profound influence on thermal transport \cite{Yamada2024} along with short-range order \cite{Drucker2023} due to thermal fluctuations above the ordering temperature. While the RKKY mechanism in these semimetals is a natural explanation, superexchange interactions mediated by the neighboring aluminum and silicon ligands can also play an important role as discussed in recent theoretical work on RAlSi compounds \cite{Bouaziz2024}. Here, DFT calculations on GdAlSi and the isostructural series predict predominantly antiferromagnetic exchange interactions without significant sign changes. In addition, isostructural SmAlSi features weak nesting around the ordering wavevector as found in $\chi''(\vb{k}_\mathrm{mag})$, but was shown to be incompatible with a magnetic instability due to the lack of divergence in $\chi'(\vb{k}_\mathrm{mag})$ \cite{Zhang2023}. 

While neutrons cannot easily probe itinerant Weyl fermions, they scatter strongly from $4f$ electrons that may interact indirectly with other $4f$ electrons through the Weyl fermions. In this work, we measured and thoroughly modeled the magnetic excitation spectrum in NdAlSi to understand the underlying interactions. We use the random-phase approximation (RPA) \cite{Jensen1991, Buyers1975, Riberolles2023, Boothroyd2020, Wen2017, Yuan2020, Nikitin2021} to model the series of weakly-dispersive, gapped, and mean-field-renormalized crystal field excitons. While neutron scattering alone cannot determine the mechanism behind the interactions, the extracted exchange constants feature sign-changing modulations consistent with expectations for the RKKY mechanism. We also show that the helical ground state in NdAlSi is stabilized by short-ranged, anisotropic Dzyaloshinskii-Moriya (DM) interactions with orthorhombic symmetry \cite{Hoffmann2017}. Previously predicted to stabilize antiskyrmions, this may be one of the first reports of a ground state spin texture arising from such low-symmetry anisotropic DM interactions in a bulk crystal, and explains the exotic magnetism arising in isostructural materials. Our work serves as the first comprehensive experimental study of magnetic excitations and interactions in a rare-earth-based Weyl semimetal, setting the stage for study of isostructural Weyl semimetals, and demonstrates the vital role of crystalline symmetry in understanding the role of Weyl fermions on helical magnetic order. We provide a first detailed view of magnetic interactions in a Weyl semimetal and an opportunity to benchmark ab-initio theories of superexchange and Weyl RKKY interactions. 

\section{Results} 
\subsection{Inelastic neutron scattering data} 
We probe the inelastic neutron scattering (INS) spectrum with various incident neutron energy $E_i$, temperatures $T$, and momentum transfers $\vb{Q} = \vb{k}_i-\vb{k}_f$. Throughout, we work in reciprocal lattice units (r.l.u.) denoted $(hkl)$ for momentum transfers $\vb{Q} = h\vb{a}^* + k\vb{b}^* + l\vb{c}^*$ where $\vb{a}^*,\vb{b}^*,\vb{c}^*$ are the mutually orthogonal reciprocal lattice vectors of the conventional unit cell. The intensity represents the differential scattering cross section $S(\vb{Q},\omega) = (k_i/k_f) \dd^2 {\sigma}/ \dd \Omega \dd E_f$, expressed throughout in absolute units, b/Sr/meV/4 Nd (see Methods). 

\begin{figure}
    \centering
    \includegraphics[width=\linewidth]{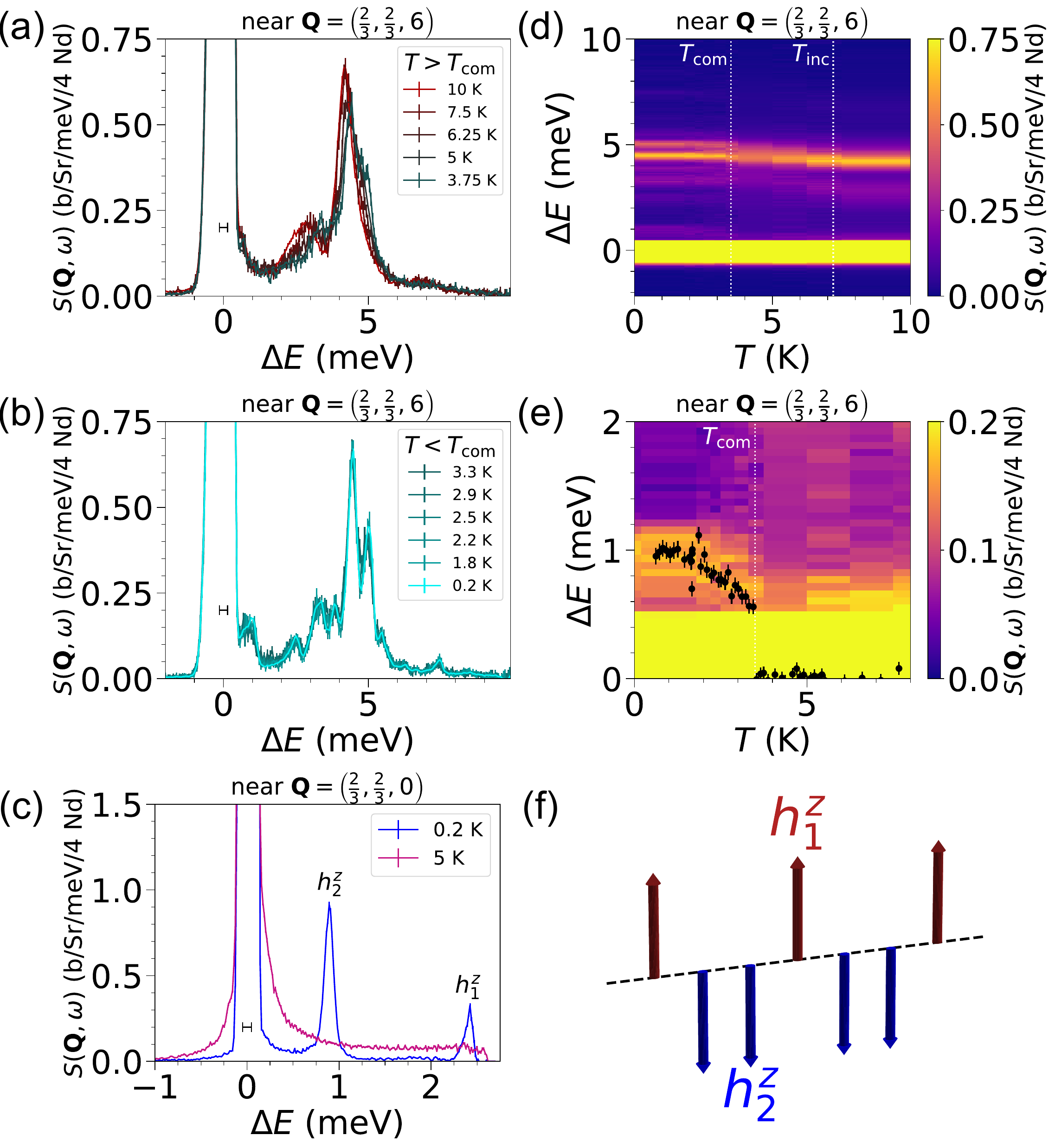}
    \caption{Temperature dependent of the inelastic magnetic neutron scattering from NdAlSi for $\vb{Q}$ near $(\tfrac{2}{3} \tfrac{2}{3} 6)$ and $(\tfrac{2}{3} \tfrac{2}{3} 0)$. (a) Spectrum above and (b) below the commensurate ordering temperature $T_\mathrm{com}\approx 3.5$ K for $E_i=11$ meV, integrated over the ranges $l\in (-6.5, -5.5), hh\in (-0.9, -0.4), k\bar{k}\in (-0.25, 0.25)$ (c) High-resolution spectra for $E_i=3$ meV above and below $T_\mathrm{com}$ showing a drastic shift of spectral weight from quasielastic to inelastic scattering, integrated over $l\in (-0.2, 0.2), hh\in (-0.7, -0.5), k\bar{k}\in (-0.15, 0.15)$. Horizontal bars in (a-c) indicate the resolution (full width at half maximum) at the elastic line. (d) Color map of the temperature dependence at $\vb{Q}=(\tfrac{2}{3} \tfrac{2}{3} 6)$ showing the splitting and sharpening of modes upon cooling into the commensurate phase. (e) Temperature dependence of the low-energy mode $\omega_1$ compared with the intensity (squared order parameter) of the $(101)$ ferrimagnetic Bragg peak (see Methods). (f) Sketch of the ordered moments in the commensurate state which experience distinct exchange fields.} 
    \label{fig:TDep}  
\end{figure}

\begin{figure*}
  \centering  \includegraphics[width=\textwidth]{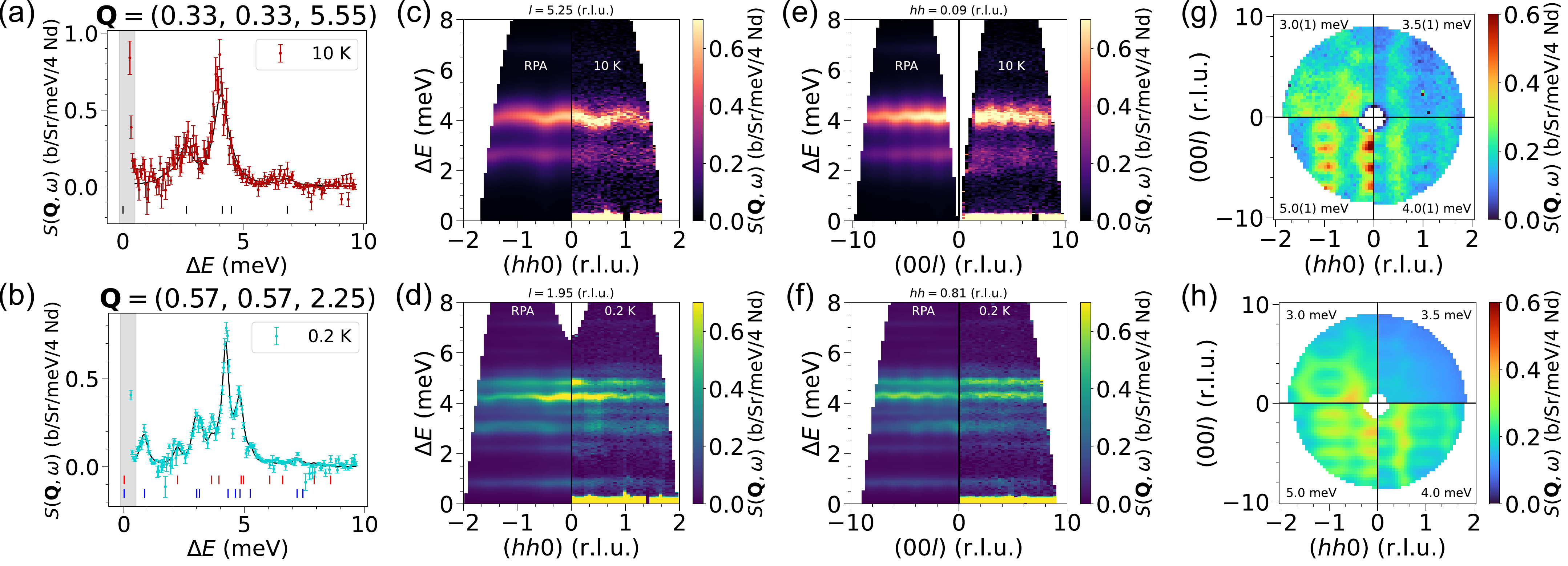}
  \caption{RPA model of the magnetic excitations. Throughout, we select general positions in momentum space, that are not necessarily of high-symmetry, to highlight the harmonics. All data were fit above 0.5 meV due to contributions from the elastic line. (a),(b) The models of the inelastic spectra at particular $\vb{Q}$-points at 10 K and 0.2 K, respectively. Data are scattered points, and black lines are the RPA calculation. The ticks represent the energy levels of the noninteracting CEF and mean-field Hamiltonian, respectively. (c),(d) The dispersion for walks along the $(hh0)$ direction at fixed $l$ for 10 K and 0.2 K respectively. (e),(f) Same for walks along $(00l)$ for fixed $hh$. In both cases, the RPA calculation is on the left (negative values), and the data are shown on the right (positive values). (g) Momentum-space $(hhl)$ maps at constant energy transfers, and (h) the corresponding simulated values.}  \label{fig:RPA_Fitting} 
\end{figure*}

First, we focus on the spectrum at isolated points in momentum space, to track the temperature dependence of the modes as they are modified by the magnetic orders. We probe the spectrum with $E_i=11$~meV neutrons at 10 K in the paramagnetic state as shown in Fig.~\ref{fig:TDep}(a). Due to the orthorhombic crystal electric field from the Al and Si ligands, the multiplet of total angular momentum $J=9/2$ is expected to split into five Kramers doublets as the irreducible representations $5\Gamma_5$ \cite{Runciman1956, Walter1984, Koster1963}. If the crystal field scheme has a first excited state well above $10$ K $\approx 0.86$ meV, then the observed modes correspond to transitions from the ground state to distinct crystal field states without significant contribution from inter-level transitions. Thus, we expect the low-T INS spectrum will consist of four transitions from the ground state to excited states. Indeed, we clearly resolve three different modes, and by fitting the crystal field scheme, we find that two of the four modes are nearly degenerate in energy (shown in Fig.~\ref{fig:RPA_Fitting}(a)). 

As we cool below $T_\mathrm{incom}\approx 7.2$ K and enter the incommensurate state, two main changes occur to the spectrum. First, the crystal field modes harden, and also significantly broaden (Fig.~\ref{fig:TDep}(b,d)). Furthermore, using $E_i = 3$ meV incident energy neutrons, which provides an improved elastic energy resolution of 0.098 meV, we find pronounced quasielastic scattering that extends above 2 meV as shown in Fig.~\ref{fig:TDep}(c). We note that such quasielastic scattering is not likely to arise from heavy-fermion physics as in other rare-earth intermetallics since DFT calculations of NdAlSi predict $4f$ bands to be located approximately 1 eV above the Fermi level \cite{Gaudet2021}. The source of this low-energy scattering becomes clear when we further cool below $T_\mathrm{com}$. The commensurate state arises to satisfy the constant-length moment constraint. Neglecting for the moment the small transverse fields arising from spin canting, the main effect of the commensurate order is the development of a periodic structure and two distinct static exchange fields at each site as a result of the $udd$ ferrimagnetic magnetic order (\ref{fig:TDep}(f)). These exchange fields act as Zeeman fields that split the Kramers-degenerate crystal field states, causing the modes to shift in energy and change in intensity. Concomitantly, the quasielastic peak is replaced with gapped, low-lying mean-field states near $\omega_1 \approx 0.8$ meV and $\omega_2 \approx 2.3$ meV. Lastly, there is an overall drastic reduction of the lifetime broadening in the ordered state compared to the incommensurate and the paramagnetic states. In particular, the intrinsic broadening $\eta$ derived from the fits to the neutron spectra (see Methods) are refined as 0.12(2) meV in the commensurate state and 0.31(6) meV in the paramagnetic state. 

The nature of the incommensurate versus commensurate order is critical in explaining the changes in spectrum. In the incommensurate state with wavevector $\vb{k} =(2/3+\delta, 2/3+\delta, 0)$, the spin structure is a long-wavelength ($a/\delta \approx 420$ \AA) amplitude-modulated spiral $m^z(\vb{r}) \sim m_0^z \cos (\vb{k}\cdot \vb{r})$ with net zero magnetization $\expval{m^z(\vb{r})}$. Although there is a distinct exchange field $h(\vb{r}_{id})$ on each site, the sample-averaged exchange field vanishes. (This would not be true for a simple commensurate antiferromagnet for example, where the distribution of fields would be bimodal at some values $\pm h_0$.) Therefore, the quasielastic scattering may reflect a mean-field distribution function $P(\{\vb{h}(\vb{r})\})$. To a first approximation we can treat this distribution as a Gaussian of a scalar field $|\vb{h}(\vb{r})|$ with mean zero and standard deviation $\Delta h$, arising from the distribution of mean fields \cite{Wen2017}. Once the magnetic structure locks into a commensurate lattice, the exchange field becomes periodic in space and bimodal with approximately two distinct exchange fields $h_1^z$ and $h_2^z$ due to the distinct nature of the majority and minority sites of the $udd$ magnetic structure. The exchange fields are themselves proportional to the spin expectation value, which increases abruptly as the ferromagnetic component onsets. Thus, the critical behavior of the first excited state is $\omega_1(T) \propto \expval{J(T)}$. Indeed, the evolution of $\omega_1(T)$ is quite abrupt near $T_\mathrm{com}$, similar to the onset of intensity at the ferromagnetic $(101)$ Bragg peak (Fig.~\ref{fig:TDep}(e)). The bi-modal exchange field distribution splits the five doublets into ten singlets with the amount of splitting depending on the site index. 

Similar to a tight-binding model for a crystalline insulator, crystal field levels can have dispersion which arise from the exchange interactions between localized spins in the lattice. Neutrons excite a site in the ground state to an excited state, which then propagates through the lattice. These excitations are therefore known as crystal field excitons, and have been found in numerous transition metal and rare-earth metal based magnets \cite{Birgeneau1971, McEwen1991, Sarte2019, Yuan2020, Gaudet2023, Gao2023}. 

Before we detail the RPA theory, let us examine the raw data and preview the corresponding RPA model. In the following cuts and slices, we show the best fits to the data for both 10 K and 0.2 K spectra, highlighting the overall agreement of the spectrum intensity, and that all modes are accounted for in our model. We note that for some modes the RPA model suffers from discrepancies in intensity relative to the experimental data, namely the high-energy modes along the $(00l)$ direction. First, Fig.~\ref{fig:RPA_Fitting}(a,b) show fits to the RPA model at different momenta for the paramagnetic and commensurate states respectively. Next, we see slight dispersion in the $(hh0)$ in Fig~\ref{fig:RPA_Fitting}(c,d), and $(00l)$ in Fig~\ref{fig:RPA_Fitting}(e,f), for both the 10 K and 0.2 K data. The dispersion is small, with a bandwidth (peak to trough) less than 0.3 meV, on top of the otherwise well-separated modes. We also note that the linewidth of some modes can slightly vary with momentum; in particular at 10 K, the low-energy 2.5 meV mode is significantly broader than the other modes. 

In Fig.~\ref{fig:RPA_Fitting}(g,h) we plot the inelastic spectrum at 0.2 K in constant-energy slices and the RPA simulation, to observe the structure of the excitations as we vary momentum transfer. We focus on the $(hhl)$ scattering plane for which there is the most coverage. The momentum space structure is rich, although there are similar structures in different bands. The slices show spots or ring-like features near integral positions, and ladder-like features that oscillate between $(1/3, 1/3, l)$ and $(2/3, 2/3, l)$ in the scattering plane. Higher momentum-space resolution data are shown in Fig.~\ref{fig:SI_MACS_data}. Such structure arises from the exchange tensor $\mathcal{J}(\vb{q})$ and the matrix elements that couple the mean-field states, as we will now discuss.

\subsection{Crystal field and exchange parameters} 
\begin{figure*}
  \centering  \includegraphics[width=\textwidth]{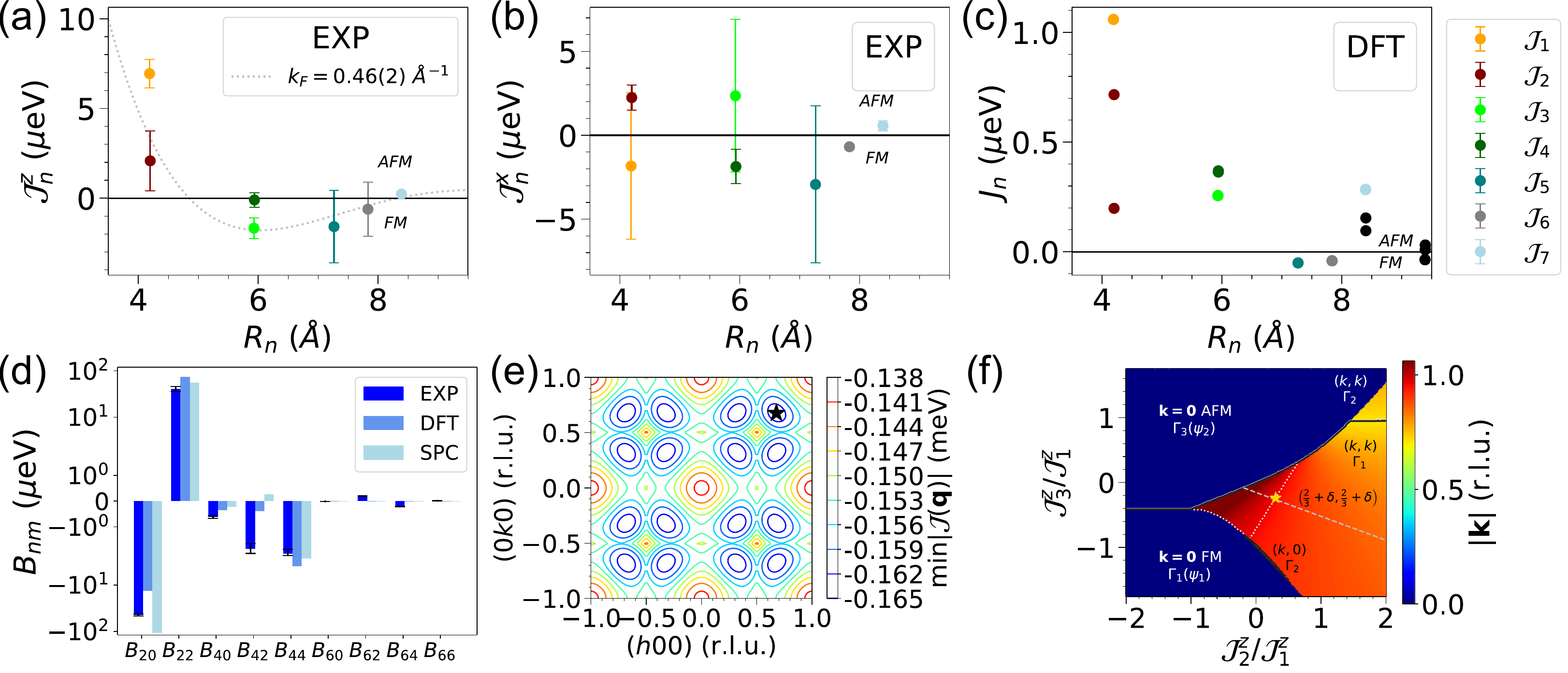}
    \caption{ Results of the fits of the Hamiltonian parameters. (a),(b) Experimental XXZ exchange parameters (EXP) for $\mathcal{J}_n^z$ and $\mathcal{J}_n^x$, respectively, as a function of near-neighbor distance. The dashed line in (a) represents the isotropic RKKY fit (see Discussion). (c) Comparison with the isotropic DFT calculation of the Gd analogue in the lattice constant of NdAlSi scaled to the Nd magnetic moment \cite{Bouaziz2024}. (d) Stevens operator coefficients $B_{nm}$ from the fits (EXP), first-principles (DFT), and the screened point charge (SPC) model (see SI). The $n\geq 6$ parameters are substantially underestimated in the DFT and SPC compared to the experimental fits. (e) Minimum energy eigenvalue calculated from the Luttinger-Tisza model, showing minima near $\vb{k}_\mathrm{mag}$ (black star) and symmetry-equivalent wavevectors. (f) Phase diagram for the fitted parameters using Luttinger-Tisza. The white dotted line is the contour for constant $|\vb{k}| = |\vb{k}_\mathrm{mag}|$, while the silver dashed line represents a constant mean-field ratio $h_2^z / h_1^z$ consistent with the data. The gold star is the parameter set of best fit.} 
    \label{fig:Exchange_CEF_RPA} 
\end{figure*}

We extract the crystal field and exchange parameters (the Hamiltonian parameters) by fits to the full $(hhl,\omega)$ spectrum of inelastic magnetic neutron scattering. The scattering intensity is calculated for given Hamiltonian parameters through the RPA, which conveniently separates the mean-field model from fluctuations beyond the mean-field scheme: 
 \begin{small} \begin{align} 
     H &= \sum_{id} H_{\mathrm{CEF}}(i,d) + \sum_{\mu\nu, id, jd'} \mathcal{J}_{id,jd'}^{\mu\nu} \hat{J}^\mu_{id} \hat{J}^\nu_{jd'} \\
     &= \sum_{id} H_{\mathrm{mf}}(i,d) + \sum_{\mu\nu, id, jd'} \mathcal{J}_{id,jd'}^{\mu\nu} \hat{J}^\mu_{id} \left( \hat{J}^\nu_{jd'} - 2\expval{\hat{J}^\nu_{jd'}} \right) 
 \end{align} \end{small} 
Here $H_{\mathrm{CEF}}$ is the crystal-field Hamiltonian from the local environment, with the single-site mean-field Hamiltonian $H_\mathrm{mf} = H_{\mathrm{CEF}} + \vb{h}_{id} \cdot \vb{\hat{J}}_{id}$, while $\mathcal{J}_{id,jd'}^{\mu\nu}$ is the exchange interaction tensor (energy per spin) with Cartesian components $\mu\nu$ and between sites $id$ and $jd'$, where $i,j$ are Bravais indices and $d,d'$ denote the basis atoms in the first unit cell; and $h_{id}^\mu = \sum_{\nu d'} 2 \mathcal{J}_{idjd'}^{\mu\nu} \expval{J^\nu_{jd'}}$ is the mean exchange field. The single-ion crystal field Hamiltonian $H_{\mathrm{CEF}}(i,d)$ is site-dependent due to nonsymmorphic four-fold rotation and mirror symmetries relating the sites $\vb{r}_1 = (0,0,0)$ and $\vb{r}_2 = (0, 1/2, 1/4)$. 

Our fitting routine is entirely self-consistent. The Hamiltonian parameters are allowed to freely refine, and a given set of these parameters will select a certain ground state spin configuration $\{ J_{id}^\mu \}$ which we find using Luttinger-Tisza. From this, we obtain a set of mean fields $\{ h_{id}^\mu \}$ and a mean-field Hamiltonian. From these, we can find the single-site non-interacting Greens function, and by calculating the exchange matrix, we obtain the interacting Greens function using the random-phase approximation, thereby producing a model for the spectrum of the excitons. Due to the large amount of fitting parameters, convergence requires good initial parameters. In the SI we describe ways of finding adequate initial crystal field parameters to inform the fitting routine. 

We use an XXZ Hamiltonian with DM interaction to fully account for the mean-field modes and the excitation bandwidth (see Methods). The results of the fits to the XXZ symmetric exchange parameters as a function of near-neighbor distance are shown in Fig.~\ref{fig:Exchange_CEF_RPA}(a,b). In the $\mathcal{J}^z$ channel, we find antiferromagnetic first and second near neighbors, followed by predominantly ferromagnetic exchange interactions for third and further neighbors, although we note that the error bars for several parameters have overlap with zero. Such predominantly ferromagnetic interactions, and the sign-changing oscillations between antiferromagnetic and ferromagnetic interactions, are consistent with RKKY-type exchange. In the $\mathcal{J}^x$ channel, we note that $\mathcal{J}_1^x, \mathcal{J}_3^x, \mathcal{J}_5^x$ are highly correlated parameters; as seen in Eq.\ref{eq:MinimumWavevector}, these contribute to the $\cos \pi H$ and $\cos 3\pi H$ harmonics that affect the dispersion. Therefore separately they suffer from substantial spread, but their linear combination is more well-constrained (see SI). We next directly compare our results to those predicted from first principle calculations \cite{Bouaziz2024} shown in Fig.~\ref{fig:Exchange_CEF_RPA}(c). Here, the electronic structure calculations are projected into spin-spin interactions assuming a Heisenberg model. In this comparison we find important similarities and differences. Firstly, we notice that while the overall exchange strength is comparable, the magnitude of the DFT calculation is smaller by up to an order of magnitude. Furthermore, the length-scale of the decay of the interaction strength is similar, decaying within about $2-3a$. We also notice anisotropic interactions for $\mathcal{J}_2$, which our neutron experiment is not sensitive to (see Methods). Next, the first two exchange constants $\mathcal{J}_1^z$ and $\mathcal{J}_2^z$ are antiferromagnetic, in line with the observed spin structure which has antiferromagnetically-aligned spins on these particular near-neighbor pairs. However, we find a large discrepancy in the third near-neighbor interaction; our model has a best fit for sign changes in $\mathcal{J}_3^z$ relative to $\mathcal{J}_1^z$ and $\mathcal{J}_2^z$. We attribute this to the fact that in Ref \cite{Bouaziz2024} and correspondingly in Fig.~\ref{fig:Exchange_CEF_RPA}(c) that the interaction has been carried out with the Gd analogue of NdAlSi in the lattice constant of NdAlSi. Furthermore, the majority of interactions are antiferromagnetic. In the superexchange mechanism, the Heisenberg exchange interaction is proportional to $\sim t_{ij}^4/\Delta_{ij}^3$ where $t$ is the hopping amplitude and $\Delta$ is the charge transfer gap, and is predominantly antiferromagnetic, although the signs depend critically on bond angles and the ligands involved \cite{Shanavas2016}. Whereas, in the RKKY mechanism, the interaction is proportional to $\sim (\cos 2k_F r_{ij})/r_{ij}^3$ and is thus sign-changing \cite{Ruderman1954, Scheie2022}, which is consistent with the exchange parameters inferred from the neutron data.

Fig.~\ref{fig:Exchange_CEF_RPA}(d) compares the crystal field parameters inferred from the experiment (EXP) with those obtained from the projector-augmented wave formalism of density functional theory (DFT), and from a screened point charge (SPC) model. The DFT and SPC models predict predict $B_{20}<0$ in agreement with the easy-axis ground state. Likewise, there is good overall agreement with the signs for most parameters among the three models. The largest difference occurs in the $n\geq 6$ angular momentum sector, which are drastically underestimated relative to the experimental values by the DFT and SPC. Apparently these theories underestimate contributions to the spatial variations of the CEF from higher spherical harmonics. As a result of this, the energy range for the CEF excitations is significantly underestimated by these theories (Fig.~\ref{fig:SI_RPA_CEF_comparison}). 

For the optimal CEF and exchange parameters obtained by fitting the inelastic neutron scattering data, let us examine the stability of the ferrimagnetic phase. In Fig.~\ref{fig:Exchange_CEF_RPA}(e) we plot the minimum energy eigenvalue of the Luttinger-Tisza energy matrix (see Methods) as a function of wavevector in the $(hk0)$ plane. We find distinct minima near $(2/3+\delta, 2/3+\delta, 0)$ and their symmetry-equivalent positions. This is to be expected, as this calculation is done in the paramagnetic state and must follow the paramagnetic space group symmetry. We identify saddle points near $(h00)$ or $(0k0)$, and we note that there is a local maximum at $\vb{k}=\vb{0}$. This indicates that the $\sim (2/3+\delta, 2/3+\delta, 0)$ wavevector is a stable minimum of the magnetic structure, and that there are no competing phases within the fixed subspace of parameters. The spin configurations and their irreducible representations predicted by this method are discussed in more detail in \ref{ssec:SpinStructure}. 

We next study the magnetic phase diagram versus exchange parameters as inferred from the Luttinger-Tisza method. By varying only $\mathcal{J}_2^z/\mathcal{J}^z_1$ and $\mathcal{J}_3^z/\mathcal{J}_1^z$ (keeping $\mathcal{J}_1^z > 0$ fixed) we can explore nearby phases in parameter space, shown in Fig.~\ref{fig:Exchange_CEF_RPA}(f). We see that the $(kk0)$ phase (with irreducible representation $\Gamma_1$) is stable over a wide range of parameter space, including for both positive and negative values of $\mathcal{J}_3^z/\mathcal{J}_1^z$. The parameters from DFT predict $\vb{k} \approx (0.72, 0.72, 0)$, in reasonable agreement with the observed incommensurate wavevector $\vb{k}_\mathrm{mag} \approx (0.677, 0.677, 0)$ to which the best-fit exchange parameters have been optimized. 

Fig.~\ref{fig:Exchange_CEF_RPA}(f) shows the ratio of mean fields on the up and down spin sites $h_2^z/h_1^z$ as calculated in Eq.~(\ref{eq:MeanFieldSimplified}) as a function of the ratios $\mathcal{J}_2^z/\mathcal{J}_1^z$ and $\mathcal{J}_3^z/\mathcal{J}_1^z$. While the $(2/3+\delta,2/3+\delta, 0)$ can be stabilized for $\mathcal{J}_3^z/\mathcal{J}_1^z>0$, a negative mean-field ratio that is consistent with the CEF spectrum requires a negative $\mathcal{J}_3^z/\mathcal{J}_1^z$, which is also required to stabilize an ordered structure with wave vector consistent with the observed wave vector very near $(\tfrac{2}{3} \tfrac{2}{3} 0)$. Our phase diagram also reveals nearby stable phases like $(kk0)$ in irreducible representation $\Gamma_2$, along with $(k00)$ phases and purely $\vb{k}=\vb{0}$ ferromagnetic $(\Gamma_1)$ or antiferromagnetic $(\Gamma_3)$ phases. All such irreducible representations for the various possible ordering wavevectors are consistent with group theory, and correspond to phases seen in other $R$Al$X$ compounds, demonstrating how our model can explain a variety of magnetic phases in these systems (with the caveat that the magnetic anisotropy can drastically vary between materials). Overall we find that there is a wide region of stability for $(kk0)$ phases near $k\sim 2/3$, and our model constrains the exchange parameters for NdAlSi through the mean field ratios and the experimentally observed magnetic wave vector. Given the extended stability range, it may require large concentration of dopants or pressures to push the magnetic state of NdAlSi into nearby phases. 

\subsection{Dzyaloshinskii-Moriya Interaction} 
While the uniaxial albeit symmetric exchange parameters discussed previously give rise to the $udd$ ferrimagnetic structure, the antisymmetric Dzyaloshinskii-Moriya (DM) interaction is a necessary ingredient to explain the canted spins in NdAlSi. The DM interaction is described with a vector quantity $\vb{D}$, in terms of which the energy is $\vb{D}_{ij}\cdot (\vb{J}_i \times \vb{J}_j)$. Thus, a collinear magnet can lower its energy if the spins slightly cant. The DM interaction generically arises as a result of spin-orbit coupling between spins that are not related to each other through inversion symmetry. Since the crystal structure of NdAlSi has no point of inversion, the DM interaction is allowed for all spin pairs. The direction and structure of $\vb{D}_{ij}$ are dictated by the symmetries of the bond between the two sites as expressed by the Moriya rules \cite{Moriya1960}. Like the symmetric exchange interaction, the strength $|\vb{D}_{ij}|$ of the DM interaction varies with the bond distance, and can even have similar oscillatory behavior \cite{Nikolic2021, Chang2015}. Therefore, it may be necessary to consider DM interactions beyond the first nearest neighbor. 

We begin by enumerating the DM interactions that are symmetry-allowed in the paramagnetic state in order of increasing near-neighbor distance. We use the method of Luttinger-Tisza to determine the spin structure for a given DM interaction, for comparison with the observed ground state. In the following discussion, we make use of the language of the spiralization tensor \cite{Hoffmann2017} to succinctly express the DM vector for particular bonds. The spiralization tensor $\mathcal{D}$ arises from a microscopic field-theoretical approach and is crucial in systems with low-symmetry interactions, expressing the real-space DM vector $\vb{D}_{ij}$ as a general matrix operation $\vb{D}_{ij} = \mathcal{D} \vu{r}_{ij}$ where $\vu{r}_{ij} = (\vb{r}_j - \vb{r}_i)/|\vb{r}_j - \vb{r}_i|$ is the unit vector between spins. The precise form of the tensor $\mathcal{D}$ will depend on the symmetry of the particular interactions \cite{Ga2022}. While the Moriya rules dictate the general form of the DM interaction for particular bonds, the spiralization tensor indicates how the DM vector transforms under transformation of bonds. If $G$ is a symmetry of the bonds, then $\mathcal{D}' = (\det G) G^{-1} \mathcal{D} G = \mathcal{D}$. In the following, we compare the spin orientation relative to $\vb{k}_\mathrm{com} = (\tfrac{2}{3} \tfrac{2}{3} 0)$, which was proposed to be perpendicular to the magnetic wavevector (that is, along $\vu{z}\times \vb{k}_\mathrm{mag}$) in previous studies \cite{Gaudet2021, Dhital2023, Yang2023}. 

\begin{figure}
    \centering
    \includegraphics[width=\linewidth]{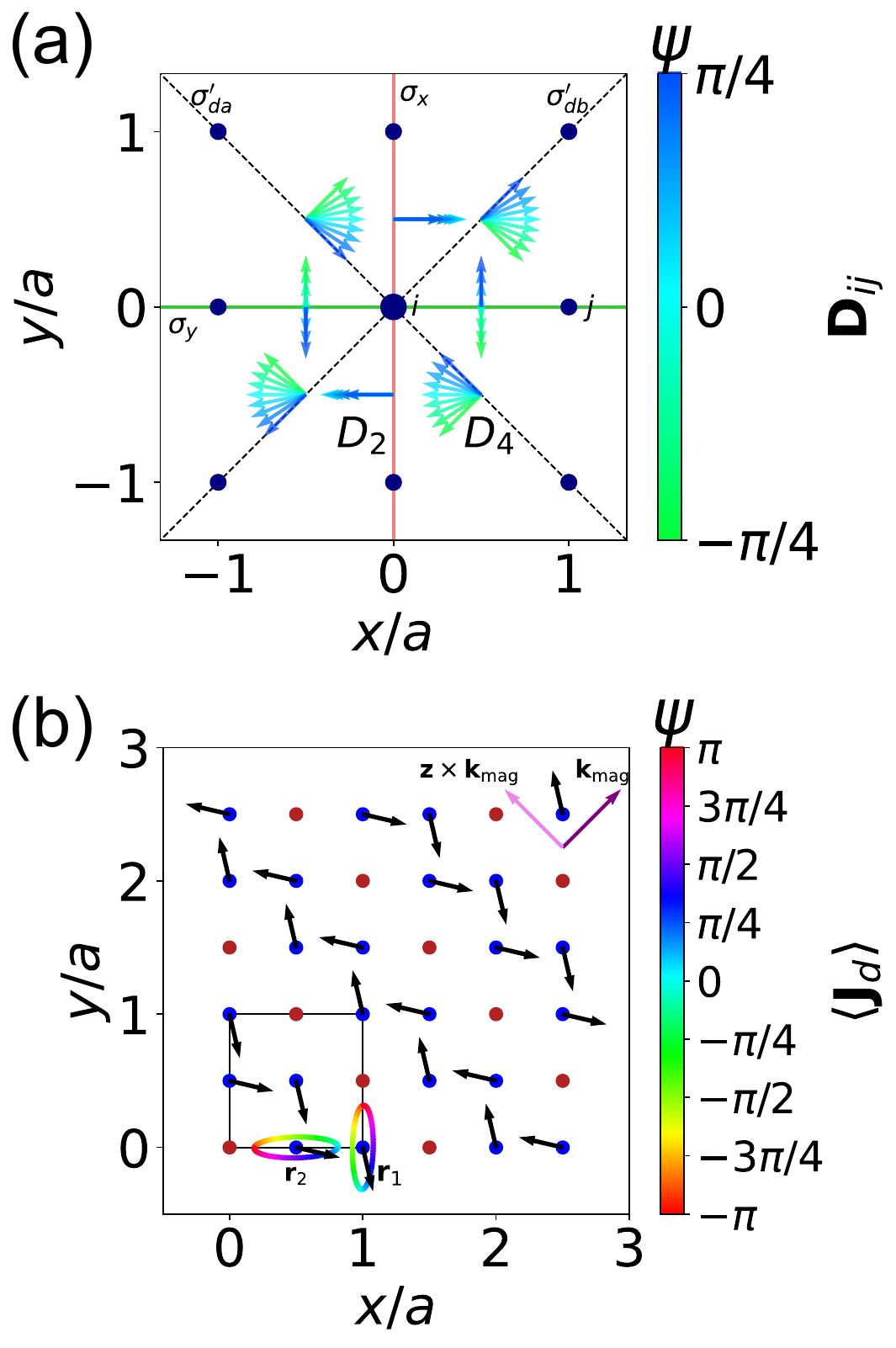}
    \caption{ Dzyaloshinskii-Moriya interaction and the resulting magnetic ground state. (a) Structure of the $z=0$ Nd atoms, symmorphic (solid lines) and nonsymmorphic (dashed lines) mirror planes, and select DM vectors for the second and fourth near-neighbors as a function of spiralization angle $\psi$. Taking a clockwise loop around the central atom results in a clockwise (anticlockwise) rotation of the DM vector given a spiralization angle of $-\pi/4$ ($\pi/4$). (b) Simulated spin structure in the magnetic unit cell determined self-consistently from the RPA refinement, with all magnetic sites projected into the $z=0$ plane, and taking $D_2 = D_x = D_y$ (equivalently, $\psi=\pi/4$) while fixing $D_4=0$. Here the in-plane spin components primarily point along the $\vu{z}\times \vb{k}_\mathrm{mag}$ direction. Up (down) spins are shown in blue (red). The in-plane canting structure sweep ellipses when changing $D_2(\psi)$. The conventional chemical unit cell is shown in the smaller black square, along with the indicated chemical basis of the primitive unit cell.} 
    \label{fig:DMI_Spins} 
\end{figure}

 \begin{enumerate} 
     \item The first near-neighbor occurs between interlayer sites $\vb{r}_1$ and $\vb{r}_2$ and lies in a mirror plane; therefore the DM vector points perpendicular to the bond. Furthermore, the bonds are related by nonsymmorphic four-fold rotation operations, which constrains the direction of the DM vector for all equivalent sites $(jd')$. We can write $\vb{D}_{id,jd'} = D \vu{z}\times \vu{r}_{id,jd'}$. However, this interaction favors a canted spin component that is parallel to $\vb{k}_\mathrm{com}$, inconsistent with experiment. 
     
     \item The second near-neighbor interaction is intralayer, occurring between atoms on the edges of the square lattice. This bond lies in a mirror plane containing the $c$-axis, and simultaneously the midpoint lies in a mirror plane perpendicular to the spins. These constrain the DM vector to point in the basal plane perpendicular to the bond. However, this bond is special, as there is no symmetry operation that takes $(i,j)$ to $(i,j')$ by a 90 degree rotation. Therefore, DM vectors for bonds along $x$ can be in general different from that of $y$, in both magnitude and direction. This result can be written as $\vb{D}_{id,jd'} = [R_d^{-1} \mathcal{D} R_d] \vu{r}_{id,jd'}$ where       
       \begin{equation} \label{eq:SpiralizationTensor}
           \mathcal{D} = \mqty(0 & D_x & 0 \\ D_y & 0 & 0 \\ 0 & 0 & 1) \equiv \mqty( 0 & |D| \cos \psi & 0 \\ |D| \sin \psi  & 0 & 0 \\ 0 & 0 & 1 ) 
       \end{equation} 
       is the spiralization tensor, with $D_x^2 + D_y^2 = |D|^2$, and 
      \begin{equation}
          R_d = \mqty(\cos \phi_d & \sin \phi_d & 0 \\ -\sin \phi_d & \cos \phi_d & 0 \\ 0 & 0 & 1)
      \end{equation} 
      is the classical rotation operation with $\phi_d = 0$ for sites $\vb{r}_1$ while $\phi_d = \pi/2$ for sites $\vb{r}_2$. For illustration, an $x$-oriented bond has $\vb{D}_{ij} = \mathcal{D} (1,0,0)^T = (0, D_y, 0)$ whereas a $y$-oriented bond has $\vb{D}_{ij} = \mathcal{D} (0,1,0)^T = (D_x, 0, 0)$. We summarize the spin canting orientation with changes in the spiralization angle $\psi$, and more importantly with the determinant $\det \mathcal{D} = -D_x D_y = -\tfrac{1}{2}|D|^2 \sin 2\psi$. Samples of real-space DM vectors for given spiralization angles $\psi$ in Fig.~\ref{fig:DMI_Spins}(a), along with their varying orientation or length. We find when $\det\mathcal{D} < 0$, the spins point nominally perpendicular to $\vb{k}_\mathrm{com}$, consistent with experiment. On the other hand, if $\det\mathcal{D} > 0$, the spins point nominally parallel to $\vb{k}_\mathrm{com}$. The distinction can be understood in how the DM vector changes orientation across the bonds. Walking along a clockwise loop around site $i$, a negative (positive) value of $\psi$ corresponding to a positive (negative) spiralization determinant will result in a clockwise (anticlockwise) rotation of the DM vector. These dichotomous orientations ultimately result in the spin canting primarily parallel or perpendicular to the magnetic wavevector. 

     \item The third nearest neighbor interaction, an interlayer interaction, has no general symmetry constraint besides the lack of inversion center, and the DM vector can be in general $\vb{D} = (D_1, D_2, D_3)$. Such a vector is difficult to constrain with three free variables. For simplicity we set it to zero here. Higher-quality data would be needed to assess this interaction. However, we consider a particular limit to compare to the prediction from field theory. Taking $\vb{D} \propto \vb{r}$, the resulting spin structure will be primarily parallel to the wavevector, inconsistent with experiment.

     \item The fourth-nearest neighbor intralayer interaction lies between the corners of the square lattice. The midpoint between the spins lies in a two-fold axis (the $c$-axis), constraining the DM vector to lie in the basal plane, $\vb{D} = (D_x, D_y, 0)$. The bonds cannot be mapped to one another by a four-fold rotation, but they do obey mirror symmetry $\sigma_x$ and $\sigma_y$. Since the DM vector is axial \cite{Thoma2021}, the mirror symmetry $\sigma_x$ will transform $(D_x, D_y, 0)$ to $(D_x, -D_y, 0)$; likewise $\sigma_y$ will transform $(D_x, D_y, 0)$ to $(-D_x, D_y, 0)$. In the language of the spiralization tensor, this can be expressed succinctly as $\vb{D}_{id,jd'} = [R_d^{-1} \mathcal{D} R_d] \vu{r}_{id,jd'}$, remarkably with the same definitions as those for the second-near-neighbor case Eq.(\ref{eq:SpiralizationTensor}). Hence the same arguments for $\det \mathcal{D}$ apply as before.      
 \end{enumerate}

In summary, in the paramagnetic state, only the second- and fourth-near neighbors accommodate DM interactions that promote the observed spin structure, and they have the identical succinct form $\vb{D}_{id,jd'} = [R_d^{-1} \mathcal{D} R_d] \vu{r}_{id,jd'}$ where the spiralization tensor $\mathcal{D}$ must have negative determinant. The origin of this anisotropic interaction is the $C_{2v}$ symmetry of the bond, which should be compared to the global $C_{4v}$ point group symmetry of the crystal and the $C_{4v}$-symmetric structure of Weyl nodes in the first Brillouin zone (Fig.~\ref{fig:CrystalStructure}(c,d)). 

It is worth noting that these  $\det \mathcal{D} < 0$ spiralization tensors are predicted to stabilize $2q$ antiskyrmion phases \cite{Hoffmann2017, Cui2022}, which have been found in bulk materials \cite{Meshcheriakova2014} and thin films \cite{Nayak2017} in applied magnetic fields. The spin texture in NdAlSi is akin to antiferromagnetic, $1q$ analog of these antisykrmion phases (see SI for more discussion). Intriguingly, $2q$ spin structures with non-trivial textures were shown to arise in isostructural Weyl semimetal CeAlGe, which feature meron-antimeron pairs, resulting in a topological Hall effect in applied magnetic fields \cite{Puphal2020}. In the SI, we show such spin configuration can be inferred from Luttinger-Tisza for a negative-determinant spiralization tensor. The lack of inversion symmetry in $R$Al$X$ compounds and the low-symmetry bonds make DM interactions ubiquitous so that nontrivial spin structures can be anticipated. 

\subsection{Spin Structure and Thermodynamics} \label{ssec:SpinStructure} 
Using a mean-field theory (MFT) treatment, we calculate the ground state spin structure that arises from the exchange interactions. The simplest configuration at $T=0$ assumes a periodic Bravais $3\times 3$ lattice corresponding to the commensurate $\vb{k}_\mathrm{com} = (\tfrac{2}{3} \tfrac{2}{3} 0)$ structure. Fig.~\ref{fig:DMI_Spins}(b) shows the calculated spin structure based on the CEF and exchange parameters that account for the INS spectrum. To provide an overview, all atoms have been projected onto the $z=0$ plane. The ferrimagnetism manifests as the $udd$ spin structure with equal-magnitude spins, while the canting arises from the DM interactions discussed in the previous section. Throughout this discussion, we describe the real-space spin canting with the spherical polar angles $(\vartheta, \varphi)$. Taking a look at the canted components projected into the $z=0$ plane, we observe that the spins are closer aligned along $\vu{z}\times \vb{k}_\mathrm{mag}$ than $\vb{k}_\mathrm{mag}$, but wobble away from its nominal direction such that $\varphi \neq -\pi/4 \approx -0.79$. 

Experimentally, this wobbling manifests in a small effect on the intensity of the magnetic Bragg peaks. Namely, the predicted structure factors are $S(\tfrac{1}{3}, \tfrac{1}{3}, 4n) \propto \tfrac{8}{3} \delta^2 (\cos \varphi - \sin \varphi)^2$ where $\delta \propto \expval{S_\perp}$ and $\varphi$ is the in-plane angle of a particular spin relative to the $x$-axis. This value is therefore a constant for all $L=4n$ (aside from the form factor). Whereas, $S(\tfrac{2}{3}, \tfrac{2}{3}, 4n) \propto \left[ \tfrac{8}{3} \hat{Q}_z^2 \delta^2 (\cos \varphi + \sin \varphi)^2 + 4(1-\hat{Q}_z^2) \eta^2 \right]$ with $\hat{Q}_z = \vu{z}\cdot \vu{Q}$, and $\delta/\eta \ll 1$ related to the spin canting, which clearly has significant contribution from the nearly-Ising spins. Our diffraction refinements with a larger set of magnetic Bragg peaks with improved statistics on the weak peaks (which are most sensitive to $\delta$ and $\varphi$) reveals a local extremum at $\varphi=-\pi/4$ and a shallow global minimum at $\varphi_0 = -1.5(1)$ (see Fig.~\ref{fig:SI_Diffraction}). In comparison, from the refinement of the Hamiltonian parameters (having taken $D_2 = D_x = D_y$) we calculate $\varphi = -1.34$, which is remarkable given that our refinement is mainly sensitive to the crystal field and magnetic excitations. More generally, for a given set of Hamiltonian parameters, Fig.~\ref{fig:DMI_Spins}(b) shows the ellipse swept out as a function of the spiralization angle $\psi$. The ellipse is preferentially aligned along $y$ ($x$) for atom 1 (2), which is a consequence of the $C_{2v}$ point group symmetry, namely there is a different energy cost for spins pointing along $x$ or $y$. Similar wobbling was observed in the easy-plane spin structure of isostructural CeAl(Ge/Si) \cite{Suzuki2019} and was equivalently explained using a $g$-tensor picture (see SI). 

\begin{figure}
    \centering
    \includegraphics[width=\linewidth]{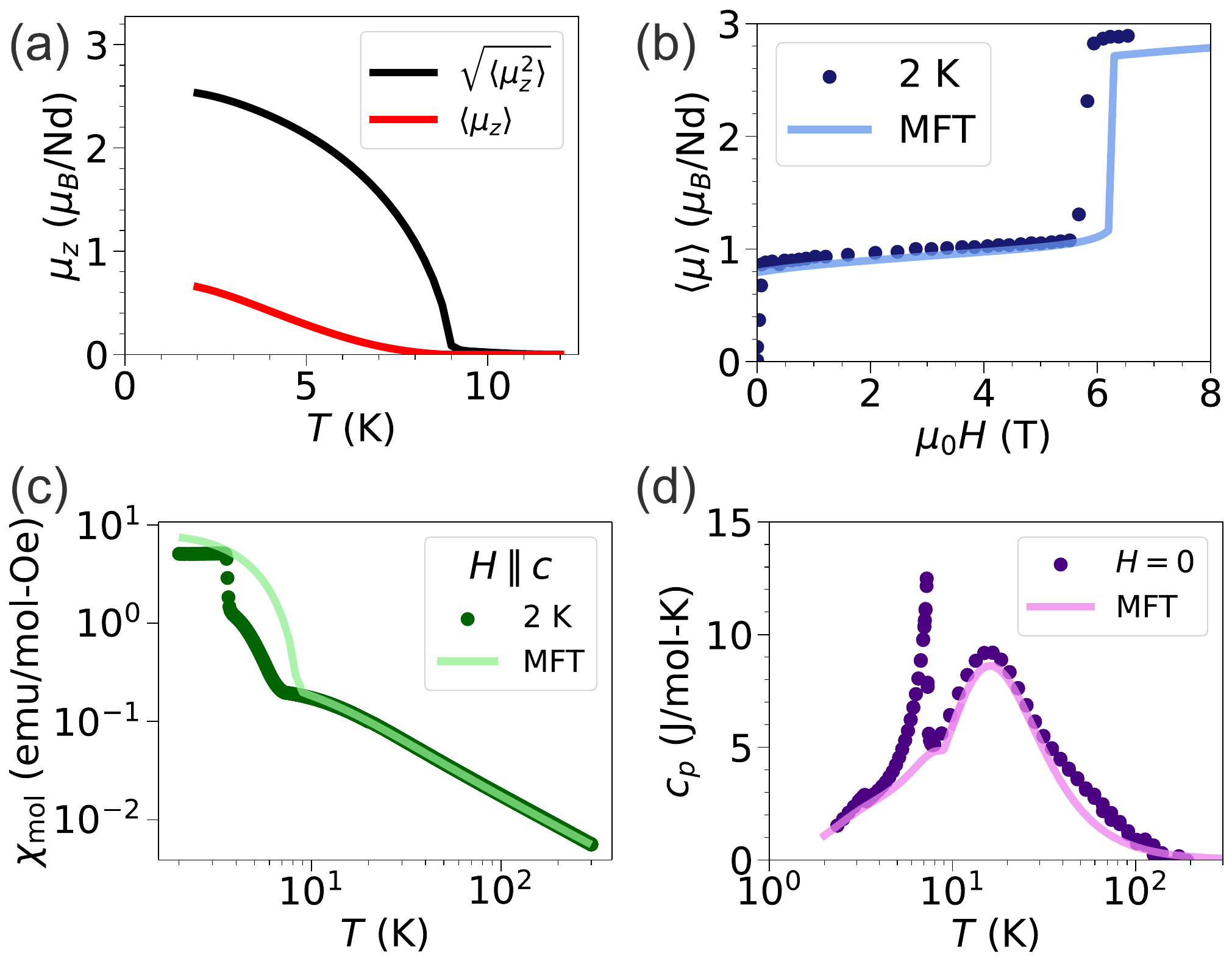}
    \caption{Thermodynamic measurements and mean-field theory (MFT) simulations. (a) Estimate of order parameters for the antiferromagnetic (black) and ferrimagnetic order (red). (b) Magnetization versus applied magnetic field for $H\parallel c$. (c) Magnetic susceptibility $\chi = M/H$ per mol Nd for $H= 500\; \mathrm{Oe}$ applied parallel to the $c$-axis, showing a transition near $T_\mathrm{inc}$ and near-perfect agreement at high temperatures. (d) Isobaric magnetic specific heat revealing the CEF Schottky and weak peaks at $T_\mathrm{inc}$ and $T_\mathrm{com}$.} 
    \label{fig:Thermodynamics} 
\end{figure}

Estimates of thermodynamic quantities associated with the Hamiltonian parameters are important tests of their predictive power. We have used a mean-field model on a commensurate lattice to determine the ground-state magnetic structure, however this is not useful in studying the changes of the spin structure in an incommensurate state. We remind that the $udd$ commensurability is a natural consequence of the constraint that the moment at each site must be constant at $T=0$, whereas in the incommensurate state the structure behaves as a long-wavelength structure akin to an amplitude-modulated spin density wave. Thus, for intermediate temperatures, we consider a model of a large $11 \times 11$ cell with open boundary conditions. The nonzero mean field $\vb{h}_{id}$ engenders a nonzero spin expectation value $\expval{\vb{J}_{id}}$, while the CEF Hamiltonian constrains the spin length through the thermal population of the mean-field renormalized eigenstates. To simulate the order parameters, we track the standard deviation $\sqrt{\expval{(J_d^z)^2}}$ to extract the mean-field incommensurate transition, and the mean $\expval{J^z_d}$ to extract information about the onset of ferrimagnetism. Within this model, we find evidence of the smooth development of ferrimagnetism and domain formation with inflection point $T_\mathrm{com} \sim 4$(1) K, above which the incommensurate state sets in, and magnetic order is destroyed near $T_\mathrm{inc}\sim 9$(1) K, as demonstrated in Fig.~\ref{fig:Thermodynamics}(a). 

We next study the field-polarized magnetic transition in the magnetization seen in Fig.~\ref{fig:Thermodynamics}(b). We self-consistently anneal the mean-field spin structure in a $3\times 3$ lattice at 2 K by sweeping various applied Zeeman fields with strength $g_J \mu_B B$, to compare with the experimental data. First, the $1/3$ magnetization plateau arises as a consequence of the $udd$ ferrimagnetic structure. In the mean-field picture, the spin lengths at 2 K are nearly their full value, so the plateau magnetization is only weakly dependent on the applied field. As we apply larger fields, there is a competition between the exchange field (from the surrounding ions) and the Zeeman energy amounting to a net energy difference $\delta H = (h_1^z - g_J \mu_B B) J_1^z$. When the applied field overcomes this difference, the minority sublattice will flip its orientation and the system will be fully spin polarized. Concomitantly, the $\omega_2 \sim |h_1| J$ mode decreases in energy until it mixes with the ground state singlet and these form the new ferromagnetic state. On the basis of such considerations alone we can predict a field-induced transition near $B_c \sim h_1^z/g_J \mu_B \sim 7.6$ T. The exact value of $B_c$ is sensitive to the crystal field scheme and exchange parameters and so is best found numerically, and in our simulation, it is remarkably close to the experimental value $B_c \approx 6$ T. 

We next study the magnetic susceptibility in weak applied magnetic fields as shown in Fig.~\ref{fig:Thermodynamics}(c). The susceptibility is sensitive to both the crystal field levels (whose population results in changes of curvature) and magnetic exchange (resulting in both Curie-Weiss behavior and changes in curvature due to magnetic order). The calculated susceptibility shows excellent agreement with the high-temperature data, and qualitatively indicates the onset of magnetic order near $T_\mathrm{inc}$ as well as a plateau below $\sim T_\mathrm{com}$. 

Lastly, specific heat capacity is another useful thermodynamic quantity that highlights phase transitions and energy level schemes. We simulate the specific heat in the mean-field model on an $11\times 11$ cell, which shows weak features near the temperature scales found in Fig.~\ref{fig:Thermodynamics}(d). The lack of available spin fluctuations in this simple mean-field model results in severe underestimates of the magnetic specific heat associated to the magnetic ordering transitions. Above the mean-field ordering temperature, we find the Schottky anomaly due to thermal population of crystal field energy levels. We remind that the above thermodynamic quantities were not directly utilized in the fitting routine, yet the model based solely upon fitting the inelastic neutron scattering data and reproducing the magnetic wave vector faithfully can account for the specific heat data versus temperature. 

\section{Discussion} 
The success of the RPA and mean-field models in modeling the spin structure and excitation spectrum of NdAlSi is owed to the large $J=9/2$ (essentially classical) total angular momentum, the significant easy-axis anisotropy, and the extended-ranged interactions. The existence of well-defined crystal field excitons facilitates modeling based on harmonic mode theories, as with several Weyl semimetals displaying clear spin waves rather than continuum scattering \cite{Rahn2017,Zhang2021}. Furthermore, the use of crystalline samples to solve the crystal field Hamiltonian allows the modeling of the weak dispersion, which cannot be resolved in polycrystals due to powder averaging. This is especially necessary in the case of NdAlSi, where the three perceptible crystal field energy levels cannot alone uniquely constrain the nine crystal field parameters.

The predictions for Weyl fermion versus free-electron RKKY mechanisms have subtle distinctions. The general forms for perfect isotropic Type-I Weyl nodes without inversion symmetry are predicted to generate Heisenberg, Ising-Kitaev, and DM interactions \cite{Nikolic2021, Chang2015}. It is then natural to assume the contribution of Heisenberg-like and Ising-like interactions can produce the effective XXZ model that we use to explain the salient features of the data. Of course, the XXZ model is not unique to Weyl metals; it has been used to explain the spin excitations of many three-dimensional magnetic systems including the honeycomb cobaltate insulators \ch{CoTiO3} and \ch{BaCo2(AsO4)2} \cite{Elliot2021, Yuan2020, Halloran2023}, the supersolid \ch{K2Co(SeO3)2} \cite{Chen2024}, and the ferromagnetic square net Dirac material \ch{CeAgSb2} \cite{Nikitin2021}, among others. It should be noted that other Dirac/Weyl semimetals, like the family (Sr/Ca)MnBi$_2$ \cite{Rahn2017} or \ch{Co3Sn2S2} \cite{Zhang2021} are well explained by isotropic Heisenberg exchange rather than XXZ exchange. Hence while there is consistency of the XXZ model and the predictions from Weyl-mediated interactions, it is not the only possible origin. We remind that the low-energy quasiparticles available to mediate the exchange interactions are small pockets of Weyl fermions, and the form of the interactions informs theories and microscopic models of RKKY interactions mediated by Weyl electrons. 

Additionally, neutron scattering is sensitive to the spin susceptibility of the rare-earth moments, and not the electronic susceptibility. While we therefore cannot directly comment on nesting and instabilities brought on by Weyl fermions, we still find hints of ``nesting" in the form of $\mathcal{J}(\vb{q})$ extracted from the INS spectrum within the Kondo approximation $\mathcal{J}(\vb{q}) \sim J_K^2 \chi(\vb{q}, 0)$, where $J_K$ is the Kondo coupling and $\chi(\vb{q},\omega)$ is the Lindhard susceptibility \cite{Fuhrman2021}. In the language of field theory of Weyl-mediated exchange, the Heisenberg exchange interactions will follow the general form $\mathcal{J}(r) \sim f(\Lambda r) e^{i\vb{K} \cdot \vb{r}}$ where $\Lambda$ is a momentum space cutoff (related to the linearity of the Weyl nodes). A small cutoff $\Lambda/|\mathbf{K}| \ll 1$ would result in a slowly-varying (i.e. constant) function $f(\Lambda r)$ with a rapidly oscillatory phase $e^{i\vb{K} \cdot \vb{r}}$, so $\mathcal{J}(\vb{q})$ would be sharply localized near $\vb{q}\approx \vb{K}$ akin to a Delta function. In contrast, we see a smooth, shallow minimum of $\mathcal{J}(\vb{q})$ at $\vb{K} = (\tfrac{2}{3} \frac{2}{3} 0)$ as shown in Fig.~\ref{fig:Exchange_CEF_RPA}(e). In the field-theory approach, this implies a larger cutoff $\Lambda/|\mathbf{K}| \sim 1$ and therefore the oscillatory behavior of the phase cannot be easily disentangled from the oscillatory behavior of $f(\Lambda r)$ given the short-ranged data. For this reason, we attempted fits to the real-space interactions $\mathcal{J}^z(r)$ with the conventional isotropic RKKY equation $R(2k_F r)$ to get a sense of the lengthscale of the wavevector \cite{Ruderman1954, Scheie2022}. The best fit value is $2k_F = 0.92(4)$ \AA$^{-1}$~ which may be compared with the reduced magnetic wavevector $\vb{k}_\mathrm{mag}'\approx (\tfrac{1}{3} \tfrac{1}{3} 0)$ with $|\vb{k}_\mathrm{mag}'|\sim 0.71$ \AA$^{-1}$~ and the paramagnetic-state separation of $|\Delta \vb{Q}_\mathrm{Weyl}|\sim 1.07 $ \AA$^{-1}$ between Weyl nodes. 

In all, the preponderance of $\mathbf{k}_\mathrm{mag}$ in the weak extrema in the spectrum from Fig.~\ref{fig:RPA_Fitting}; the weak minima in the Luttinger-Tisza minimum eigenvalue $\lambda(\vb{q})$; and the natural lengthscale of oscillations in the radial lengthscales are consistent with the electronic susceptibility \cite{Zhang2023} and first principles \cite{Bouaziz2024} calculations for $\mathcal{J}(\vb{q})$. Some notes are in order, however. First, it was noted by Bouaziz \textit{et al.} \cite{Bouaziz2024} that this weak peak in $\mathcal{J}(\vb{q})$ persists in DFT even in the centrosymmetric analog without Weyl fermions. Note however, that the centro-symmetric model would not account for the chiral spin canting associated with DM interactions. Furthermore, we find that if Weyl fermions are responsible for the significant weight on this harmonic, their effects are not as singular as might be expected for longer-ranged oscillations with more beat frequencies, and hence their effects are not easily distinguished from that of an RKKY interaction from conventional itinerant electrons.

We note that the paramagnetic crystal field modes at 10 K do not all have the same widths, even when correcting for resolution. In particular the $\sim 2.8$ meV energy mode can broaden as wavevector is varied as seen in Fig.~\ref{fig:RPA_Fitting}(c,e). This cannot be easily captured in our simple model where we take the intrinsic broadening as a constant. It is not immediately obvious why this mode is broader than the others, however in this temperature regime it was shown that spin fluctuation-enhancement of the Nernst effect was found due to the nodal character of the Fermi surface \cite{Yamada2024}, and so different crystal field modes may couple differently to the conduction electrons that results in their reduced lifetime. It may be also worth investigating, in addition to this effect, the spin excitations in isostructural Weyl semimetals like CeAlGe. This material orders into a $\vb{k}=\vb{0}$ structure (or slightly incommensurate, depending on doping levels \cite{Suzuki2019, Puphal2020}) but has evidence of short-ranged magnetism and spin fluctuations near $\vb{k}_\mathrm{mag}\sim \Delta \vb{Q}_\mathrm{Weyl}$ at temperatures above the ordering temperature \cite{Drucker2023}. It would be beneficial to probe the effect of this fluctuation by means of INS; if this effect arises generically from Weyl nodes of separation $\Delta \vb{Q}_\mathrm{Weyl}$, they may leave an imprint on certain harmonics (similar to $\cos 3\pi h$ in NdAlSi) despite not developing instability (in the form of long-range order) at this harmonic. 

We now discuss the role of the DM interactions in our work. The broken inversion symmetry can simultaneously lead to the existence of Weyl nodes and DM interactions for any bond. We refine a DM interaction on the second-near neighbor of strength $|D_2| = 2.6(3)$~$\mu$eV to get consistency with the spin structure and overall spin canting. While the DM interaction plays a significant impact in the spin structure, its effect on the excitation spectrum is subtle. Firstly, the real-space DM vectors which self-consistently produce the spin structure lie only in the basal plane, which implies that the Fourier transform of the interaction matrix $D^{\mu\nu}_{dd'}(\vb{q})$ will not have $l$-dependence. Therefore, its effect on dispersion will compete with the various harmonics from the symmetric exchange contributions and is not easily discerned. For this reason, we only attempt fits to the second-near neighbor interaction. However, the role in the energy level scheme is similar to other quantum magnets in that it will generally split energy modes due to the reduced symmetry \cite{Gaulin2004, Chen2018}.

In a field theoretical approach for local-moment interactions driven by isotropic Weyl fermions, the DM vector is predicted to be bond-parallel, and is derived from a sum over pseudoscalar chiralities and momentum-space positions of the Weyl nodes. In a more general case with tilted Weyl cones (see SI), the DM vector also has a component perpendicular to the bond direction, providing a general form $\vb{D}(\vb{r}) = D_{\vu{r}} \vu{r} + D_{\vu*{\theta}} \vu*{\theta}$, where $D_{\vu{r}}, D_{\vu*{\theta}}$ involve sums over all Weyl nodes and the tilting vectors $\vb{u}_n$. Recall the local moments (Weyl nodes) are positioned in real space (the Brillouin zone) and are subject to nonsymmorphic (symmorphic) symmetries. Therefore if only the pseudoscalar chirality and Weyl node position appear in the Hamiltonian kernel $H_\mathrm{DM}(\vb{k})$, the consequent DM interaction in the paramagnetic state will follow the symmetry of point group $C_{4v}$. Although $C_{4v}$-symmetric DM vectors can produce spin canting, they do not result in the momentum-perpendicular spin canting found in experiment. If not arising from the Weyl electrons, the shorter-ranged $C_{2v}$-symmetric anisotropic interactions that give rise to the momentum-perpendicular canting may be facilitated by superexchange interactions as found in noncentrosymmetric magnets like MnSi \cite{Moriya1960, Shanavas2016}. 

In the ferrimagnetic state, the magnetic space group $Cc'$ only retains the identity and the antiunitary nonsymmorphic mirror symmetry. This symmetry breaking dramatically reduces the number of elements in the point groups, and therefore has an effect on the Weyl nodes in addition to band folding \cite{Li2023}. The point group symmetry of the real-space bonds (Weyl nodes) is reduced from $C_{2v}$ ($C_{4v}$) to $C_1$ ($C_{1h}$). This can introduce magnetoelastic relaxations that can thus engender additional DM interactions. This includes bond-parallel interactions predicted for Weyl fermions that were not present in the paramagnetic state, and can additionally affect the spin canting. Similar behavior arises in multiferroics where the breaking of inversion symmetry by magnetic order will give rise to DM interactions that were not allowed in the paramagnetic state, and select a chiral spin structure \cite{Cabrera2009}. In principle, all possible DM interactions can compete, but it is not possible to differentiate these in our work due to their minute contribution to the excitation spectrum. Concurrently, we have seen that DFT calculations predict primarily isotropic and antiferromagnetic interactions, and have not resolved DM interactions of comparable strength to the symmetric exchange interactions. Conversely, our work points to sign-changing symmetric exchange and low-symmetry antisymmetric exchange interactions to self-consistently resolve the spin structure and its excitation spectrum. Together, these motivate further high-precision DFT calculations and ab-initio theories of Weyl-mediated exchange with the full orbital and spin degrees of freedom to pin down the predicted electronic origin and structure of the anisotropic DM interactions. 


Overall, we have quantitatively deduced the magnetic interactions present in the helical Weyl ferrimagnet NdAlSi by fitting the spectrum of crystal field excitons. We have shown the presence of sign-changing oscillations that may be consistent with coupling between itinerant electrons and localized $4f$ electrons. In light of recent studies on the strong renormalization of band structure in NdAlSi, and the coincident length scales of RKKY-type interactions with the Weyl node separation, this may make NdAlSi special in the family for its strong coupling to conduction electrons, and warrants further theoretical investigations. Furthermore, we have discussed in detail the DM interactions and find that independent of the particular near-neighbor coupling, the mechanism that best explains the observed spin texture is that of a spiralization tensor with negative determinant, which gives rise to antiskyrmion lattices or single antiskyrmions if the right parameter sets of temperature and external magnetic field is met for formation. This model is notable not only for its simplicity but also for its predictive power. It will be fruitful to compare the scattering spectrum of other rare-earth $R$AlSi materials, which may indicate the universality of harmonics at the Weyl-node separation. Lastly, given the relatively weak ($\mu$eV) interaction energy scales in NdAlSi, it may be interesting to study structurally-related materials that harbor magnetic transition metals and may have stronger quantum and correlation effects, especially in route to finding correlated topological materials, where our work may help inform experimental manifestations of coupling to Weyl fermions in these ordered states.

\section{Methods}
\subsection{Neutron scattering}
This work includes the results of four main neutron scattering experiments with crystals that were synthesized as previously reported \cite{Gaudet2021}. We performed an initial neutron diffraction experiment at the BT-4 Triple Axis Spectrometer which included the results of the magnetic $(101)$ Bragg peak in varying temperature. The magnetic peak was fit as a power law $I(T) = A_0 (1-T/T_c)^{2\beta} + I_0$, and plot in Fig.~\ref{fig:TDep}(e) by scaling $T$ to match the $T_\mathrm{com}\approx 3.5$ K. Next, we performed initial inelastic neutron scattering experiment at the CHRNS Multi-Axis Crystal Spectrometer (MACS), which qualitatively show the excitations in the $(hhl)$ plane. We used 0.44(1) mg of single crystals and glued them onto a thin aluminum plate with the fluoropolymer adhesive Cytop, and measured with $E_f = 3.7$ meV neutrons. The large step sizes in energy precluded careful quantitative fits but allowed for the observation of the complex dispersion in constant-energy slices. Some of these data are shown in SI in Fig.~\ref{fig:SI_MACS_data}. In the above data, the error bars represent one standard deviation. 

We performed comprehensive inelastic neutron scattering using the Cold Neutron Chopper Spectrometer (CNCS) at the Spallation Neutron Source, Oak Ridge National Lab (ORNL). We co-aligned 25 single crystals and glued them onto aluminum plates with minimal amounts of Cytop. The samples were aligned so that the experiment probed primarily the $(hhl)$ scattering plane, with $\pm 15^\circ$ out-of-plane coverage of the $(k\bar{k}0)$ direction. The total sample mass was 1.75(1) g with a mosaicity of 2.4$^\circ$ as determined by in-plane rocking scans. We performed measurements down to 0.2 K using a He-3 insert, well below the commensurate transition $T_\mathrm{com} \approx 3.0$ K. We used incident energies of $E_i=$ 3 meV, 11 meV, and 25 meV, the latter of which demonstrate that the modes are accounted for up to 10 meV (see SI). We measured at several intermediate temperatures up to $T=10$ K to explore the spectrum between $T_\mathrm{com}$ and $T_\mathrm{inc}$, along with in the paramagnetic state at $T=10$ K (above $T_\mathrm{incom}$). Lastly, we measured an aluminum background holder with cylindrical symmetry and whose exposed volume had nearly identical mass to the plates, with similar volume of Cytop. The reported data are expressed as $I_\mathrm{data}(\vb{x}_i) = I_\mathrm{sample}(\vb{x}_i) - I_\mathrm{empty}(\vb{x}_i)$ and with standard error $\delta I_\mathrm{data} = \sqrt{\delta I_\mathrm{sample}(\vb{x}_i)^2 + \delta I_\mathrm{empty}(\vb{x}_i)^2}$; that is, using a self-shielding factor of 1. We performed mesh scans from $\Omega=0$ to 180$^\circ$ in steps of 1$^\circ$ for 0.2 K and 10 K (at intermediate temperatures we used 2$^\circ$ steps, and for the background used 2$^\circ$ steps). The data were also symmetrized with respect to mirror symmetry operations $\sigma_x,\sigma_y,\sigma_z$ to improve statistics. Unless otherwise noted, the 11 meV data in cuts, slices and volumes were integrated over ranges $H\bar{H}\pm 0.1$ r.l.u. (due to the negligible dispersion in this range), while $HH\pm 0.05$ r.l.u. and $L\pm 0.25$ r.l.u. (amounting to roughly $\delta Q \approx \pm 0.1$ \AA$^-1$) to improve statistics, considering the aspect ratio $(2\pi\sqrt{2}/a)/(2\pi/c) \sim 5$ of the two orthogonal axes in the $(hhl)$ scattering plane. The energy axis was binned $\delta E = 0.055$ meV which may be compared to the resolution (FWHM) of 0.283 meV at the elastic line. All inelastic neutron spectra are normalized to absolute units (denoted throughout as ``abs. units''), barn/steradian-meV per conventional unit cell, by normalization to the nuclear $(004)$ Bragg peak which does not gain intensity in the magnetically ordered state and for which $|F_N(004)|^2 = 2.32$ barn/u.c.

Lastly, we performed comprehensive neutron diffraction at the DEMAND HB-3A single crystal diffractometer in the High-Flux Isotope Reactor at ORNL. We measured a 140mg single crystal nominally oriented in the $(hhl)$ scattering plane, and for which we had out-of-plane coverage between $-8^\circ$ and $40^\circ$ to access a wide array of nuclear and magnetic Bragg peaks. We used $1.005$ \AA~ neutrons for which there was negligible higher-order contamination and thus did not require the use of a PG filter. Due to the relatively large sample, there are non-negligible extinction effects that were corrected for using an empirical extinction formula \cite{Chen2020, RODRIGUEZCARVAJAL1993} $I_{\mathrm{obs}} = I_{\mathrm{th}} / \sqrt{1+10^{-3} b\lambda^3 I_{\mathrm{th}}/\sin 2\theta}$. The scaling factor and overall extinction parameter were fit then fixed from the data of the 10 K nuclear Bragg peaks. Peaks in the commensurate state were measured down to 1.5 K in an orange cryostat.

\subsection{Crystal field theory}
Throughout, we refer always to the global (lab frame) coordinate system $xyz$, where $\vu{z} = \vu{c}$ is the quantization axis. Coordinate rotations therefore are not necessary when calculating expectation values, crystal field states, or projections of Greens functions. Any changes of local coordinate systems, for example due to local crystal field environment, is performed as an active transformation of the quantum mechanical Hamiltonian. We use $\vb{J}$ to denote the quantum operator for the total angular momentum $J=9/2$ (not to be confused with the magnetic exchange parameters $\mathcal{J}$). We neglect the higher-energy $J$ multiplets in this approach, due to the maximal crystal field energy splitting of about $\Delta_\mathrm{CEF}\sim 10$~meV, compared to the $J=7/2$ multiplet at about $\Delta_{7/2}\sim 250$~meV, where admixture of states goes roughly as $\Delta_\mathrm{CEF}/\Delta_{7/2}$ \cite{Boothroyd1992}. 

Each Nd ion has $C_{2v}$ point group symmetry, with mirror planes $\sigma_x$ and $\sigma_y$ perpendicular to the quantization axis $\vu{z}\propto \vb{c}$ along which there is also a 180-degree rotation operation $C_{2z}$. Due to these mirror planes, there will be no contributions from the odd-parity Stevens coefficients $B^{(s)}_{nm}$, leaving only contributions from the even-parity coefficients $B_{nm}$. Furthermore, from group theory \cite{Walter1984} there will be in general nine independent crystal field parameters $\{B_{nm} \} = \{ B_{20}, B_{22}, B_{40}, B_{42}, B_{44}, B_{60}, B_{62}, B_{64}, B_{66} \}$. The Hamiltonian is generally written, in the Stevens operator formalism, 
 \begin{gather} 
     H_0 = \sum_{nm} B_{nm} O_{nm}(\vu{J}) 
 \end{gather} 
where $O_{nm}(\vu{J})$ are the Stevens operators which are products and linear combinations of the components of the total angular momentum $\vu{J}$. For example, $O_{20} = 3 \hat{J}_z^2 - J(J+1)$ and $O_{22} = \tfrac{1}{2} (\hat{J}_+^2 + \hat{J}_-^2) = (\hat{J}_x^2 - \hat{J}_y^2)$. \cite{Danielsen1972}. 

There are four basis atoms in the conventional unit cell that we denote as site 1 $(0,0,0)$, site 2 $(1/2,0,1/4)$, site 3 $(1/2,1/2,1/2)$ and site 4 $(0,1/2,3/4)$. Atoms $1$ and $3$, and likewise $2$ and $4$, are identical in the primitive non-orthogonal cell. Atoms 1 and 2 are related by non-symmorphic four-fold rotational ${C_{4z}^\pm}' \equiv \{ C_{4z}^\pm |\vb{t}_\pm \}$ and mirror plane $\sigma_{db}' = \{ \sigma_{db}|\vb{t}_+\}, \sigma_{da}' = \{ \sigma_{da}|\vb{t}_-\}$ symmetries, where $\vb{t}_+ = (0, 1/2, 1/4)$ and $\vb{t}_- = (1/2, 0, 3/4)$. This means their crystal field operators are related by the quantum mechanical rotation $\mathcal{R}_{001}(\pi/2)=e^{-i\hat{J}_z \pi/2}$. The CEF Hamiltonian for a general site can then be expressed as 
 \begin{equation} \label{eq:QM_Rotation_Hamiltonian}
     H_{\mathrm{CEF}}(i,d) = \mathcal{R}_d^{-1} H_0 \mathcal{R}_d 
 \end{equation} 
where $\mathcal{R}_d = e^{-i\hat{J}_z \phi_d}$ with $\phi_1 = 0$ for atom 1 and $\phi_2 = \pi/2$ for atom 2. This can also be written in terms of the nonsymmorphic mirror operation $\sigma_{db} = 2_{1\bar{1}0} \Pi$, where $2_{1\bar{1}0}$ is the two-fold rotation along the axis perpendicular to the mirror plane, and $\Pi$ is the inversion operator. As $\vb{J}$ is invariant under $\Pi$, we can equivalently write the quantum mechanical action of $\sigma_{db}$ as $\mathcal{R}_d = e^{-i\vu{J}\cdot \vu{n} (2\phi_d)}$ where $\phi_d$ is the same as before and $\vu{n}=(1,-1,0)/\sqrt{2}$. In all cases the effect of these nonsymmorphic symmetries is encoded in the eigenvectors $\ket{n}_{id}$, but in the paramagnetic state, the energy level scheme is independent of $d$. 

\subsection{Exchange Hamiltonian} 
Throughout, we follow the convention of \cite{Buyers1975}. We use an XXZ symmetric exchange interaction which can be expressed as $H_\mathrm{XXZ} = \sum_{idjd'} \left( \mathcal{J}^x_{id} [J^x_{id} J^x_{jd'} + J^y_{id} J^y_{jd'}] + \mathcal{J}^z_{idjd'} J^z_{id} J^z_{jd'} \right)$. We also include the antisymmetric exchange interaction that arises from the DM interaction $H_\mathrm{DM} = \sum_{idjd'} \vb{D}_{idjd'} \cdot \vb{J}_{id} \times \vb{J}_{jd'}$. The Fourier transform of the exchange parameters, $\mathcal{J}_{dd'}^{\mu\nu}(\vb{q})$, forms a Hermitian matrix which can be written as 
 \begin{multline}
     \mathcal{J}_{dd'}^{\mu\nu} (\vb{q}) = \sum_j \mathcal{J}^{\mu\nu}_{id,jd'} e^{-i\vb{q}\cdot (\vb{R}_{id} - \vb{R}_{jd'})} \\
     = \mqty( \mathcal{J}^{x}_{dd'}(\vb{q}) & +D^z_{dd'}(\vb{q}) & -D^{y}_{dd'}(\vb{q}) \\ -D^z_{dd'}(\vb{q}) & \mathcal{J}^{x}_{dd'}(\vb{q}) & +D^x_{dd'}(\vb{q}) \\ +D^y_{dd'}(\vb{q}) & -D^x_{dd'}(\vb{q}) & \mathcal{J}^{z}_{dd'}(\vb{q}) ) 
 \end{multline}
where 
 \begin{gather}
 J^\mu_{dd'}(\vb{q}) = \sum_{j} J^\mu_{0djd'} e^{-i\vb{q}\cdot (\vb{R}_{0d}-\vb{R}_{jd'})} \\
 D^\sigma_{dd'} (\vb{q}) = \sum_{j} D^\sigma_{0djd'} e^{-i\vb{q}\cdot (\vb{R}_{0d}-\vb{R}_{jd'})} 
 \end{gather}
are the Fourier transform of the symmetric exchange terms and DM vector, respectively, while 
 \begin{gather} 
    D^{\mu\nu}_{dd'}(\vb{q}) = \sum_\sigma \epsilon^{\sigma\mu\nu} D^\sigma_{dd'}(\vb{q}) 
 \end{gather}
 are the tensor components of the anisotropic part of the exchange Hamiltonian. The totally antisymmetric tensor $D_{idjd'}^{\mu\nu}$, made from the vector components of $\vb{D}_{idjd'}$, should not be confused with the spiralization tensor $\mathcal{D}$. 

We note a few limitations of this model. First, we index the exchange parameters by the radial distances $R_n$ as $\mathcal{J}^{\mu\mu}_{idjd'} = \mathcal{J}_n^\mu \delta_{R_n, |\vb{r}_{idjd'}|}$, and similarly for the DM interaction. However, this does not capture the anisotropy that exist in some (symmetric) exchange parameters that have the same radial distance. For example, $\mathcal{J}_2$ between sides of the square lattice has generically different values for the $x$-direction and the $y$-direction that is not captured in our model. This anisotropy would be most prominent in the $(hk0)$ scattering plane, which our experiment is not sensitive to (see SI for coverage and excitation spectrum). We note that this anisotropy is taken into account for the antisymmetric exchange in the spiralization tensor formalism. Next, while the XXZ model is one degree of complexity higher than the isotropic Heisenberg model, the general exchange matrix has several more independent parameters, as has been described in detail by Suzuki \textit{et al.} \cite{Suzuki2019}. In addition to the antisymmetric off-diagonal terms, there are symmetry-allowed symmetric off-diagonal terms, along with independent diagonal elements. We remark that our model needs at least the XXZ anisotropy to account for the bandwidth of the modes. Lastly, since the different crystal field states $\ket{n}$ have different and anisotropic probability density functions, exchange couplings between spins can likewise in principle depend on their respective crystal field state. This would add considerable complexity and free parameters that are not likely to be well-constrained by our dataset. In all, our model serves as a best estimate of the underlying strength of the exchange parameters as a function of distance that are consistent with the spectrum, and can be compared with estimates derived from DFT and the general structures predicted from field-theoretical approaches. 

We refer to \cite{Bouaziz2024} and the Supplementary Information therein on the details of the DFT+U for the crystal field and first-principles calculations of the Heisenberg exchange constants. We calculate the exchange constants by first calculating the exchange parameters derived from infinitesimal rotation of paramagnetic spins in the GdAlSi analog but with the lattice constants of NdAlSi. Note that to facilitate comparison with the convention $\sum_{ij} \mathcal{J}_{ij} \vb{J}_i \cdot \vb{J}_j$ in this work, we multiply by the factor $-\frac{1}{2} \frac{(g_J-1)^2}{J(J+1)}$ to account for the de Gennes factor and total angular momentum of Nd$^{3+}$. 

\subsection{Mean-field formalism} 
With the small interaction strengths the main perturbation to the dispersionless crystal field excitations are the renormalization of the energy level scheme due to the development of a mean field $h_{id}^\mu$ at each site; the next order perturbation is the dispersion of these levels, which we model in the RPA treatment. The effect of magnetic order is the development of nonzero expectation value of angular momentum $\expval{J^\mu_{id}}$ on every site. This produces a net mean field $h_{id}^\mu = \sum_{jd'} 2 \mathcal{J}_{idjd'}^{\mu\nu} \expval{J^\nu_{jd'}}$, where the mean field with site index ($id)$ is due to the ordered moments on its neighbors at sites indexed by $\{(jd')\}$. The factor of 2 arises from the double-counting done in writing the Hamiltonian as a sum over sites $i,j$. Note that the expectation value of the operator $\vu{J}$ is related to but should not be confused with the ordered moment. We remind that for rare-earth ions with total angular momentum $J$ following Hund's rule, the moment is $\vu{m}_{id} = g_J \mu_B \vu{J}_{id}$ whose expectation value follows the same form. Note $g_J$ is isotropic, and any anisotropy arising from the local (crystal field) environment is endowed into the wavefunctions $\ket{n_{id}}$ which can be used to calculate the total angular momentum expectation value self-consistently through 
 \begin{equation} 
     \expval{\hat{J}^\mu_{id}} = \tr (\rho_{id} J^\mu_{id}) = \sum_{n_d} \prescript{}{d}{\bra{n}} \hat{J}^{\mu} \ket{n}_d e^{-\beta \omega_{nd}}/ \sum_{n_d} e^{-\beta \omega_{nd}} 
 \end{equation} 
where $H_\mathrm{mf} \ket{n}_d = \omega_{nd} \ket{n}_d$ is the eigenvalue problem for the general single-site mean-field Hamiltonian 
 \begin{equation}
     H_\mathrm{mf} = H_\mathrm{CEF} + \vb{h}_{id} \cdot \vb{J}_{id} - g_j \mu_B \vb{B}\cdot \vb{J}_{id} 
 \end{equation} 
From the mean-field Hamiltonian we can compute thermodynamic quantities like the magnetization $\expval{M(H)}$, specific heat $C_p(T) = \dv*{\expval{E}}{T}$, and the magnetic susceptibility $\chi(T) = \expval{M(T,H)}/H$. Of course, such mean-field calculations neglect spin dynamics. 

We notice that the spin structure $\{\expval J^\nu_{jd'} \}$ and mean fields $\{ h_{id}^\mu \}$ will inform one another. If we assume a collinear ferrimagnetic $udd$ spin structure at $T=0$ and neglecting spin canting, then the net mean fields can be calculated exactly and to arbitrary distance (below shown to ninth near neighbor), 
\begin{small}  \begin{subequations} \label{eq:MeanFieldSimplified}
\begin{eqnarray}
     h_1^z \approx -8J (\mathcal{J}_1^z + \mathcal{J}_2^z - \mathcal{J}_5^z + 2 \mathcal{J}_7^z + \mathcal{J}_8^z) \sim -0.34 \; \mathrm{meV} \\
     h_2^z \approx -8J (\mathcal{J}_3^z + \tfrac{1}{2} \mathcal{J}_4^z + \mathcal{J}_5^z + \mathcal{J}_6^z + \mathcal{J}_9^z ) \sim +0.12\;\mathrm{meV} 
\end{eqnarray}
\end{subequations}  \end{small} 
Here $\mathcal{J}_n^z$ are the out-of-plane symmetric exchange components in units of meV, while $\expval{J^z_d} \approx \pm J$ is the ordered angular momentum for $J=9/2$ of Nd$^{3+}$, with $h_1^z$ referring to the field on the up site $(+J)$ and $h_2^z$ referring to the field on the down sites $(-J)$. The estimated values of $(-0.34, +0.12)$ meV are those obtained from the best fit values. Note that the site index $d$ does not appear in these equations, namely that mean fields on different sites are the same as long as they correspond to up or down spins. They also demonstrate that at least three near neighbors are required to explain this spin structure, else only one nonzero mean field would arise. The lack of $\mathcal{J}_1^z$ and $\mathcal{J}_2^z$ contributing to $h_2^z$ is due to cancellation of summing equal amounts of $+J$ and $-J$ neighbors. For fixed values of these two mean fields, clearly the dimensionality of the large number of exchange constants cannot be reduced by more than two, but such constraint helps to inform their estimated values. Lastly, with the form of the Hamiltonian described previously, the mean field and its associated ordered spin must have opposite signs so as to minimize the ground-state energy. Consequently, $h_1^z$ and $h_2^z$ must differ in sign because they represent the mean fields of moments with different signs. 

From Eq.\ref{eq:MeanFieldSimplified}(a,b), it is clear that there must be sign changes in the exchange parameters to allow for $h_1^z$ and $h_2^z$ to have opposite sign. Namely, in this approximation $(\mathcal{J}_1^z + \mathcal{J}_2^z - \mathcal{J}_5^z + 2\mathcal{J}_7^z + \mathcal{J}_8^z + \cdots )$ must be positive and $(\mathcal{J}_3^z + \tfrac{1}{2} \mathcal{J}_4^z + \mathcal{J}_5^z + \mathcal{J}_6^z + \mathcal{J}_9^z + \cdots)$ must be negative. Not only is this a fortuitous result that cannot be easily predicted from first principles, it also gives further indication that sign-changing interactions are needed to stabilize the observed magnetic order in NdAlSi. 


\subsection{Random-phase approximation}
Proceeding from the mean-field formalism, we separate the Hamiltonian into the combination of single-site mean-field term and fluctuations away from the mean-field, 
 \begin{multline}
     H = \sum_{id} \left( H_{\mathrm{CEF}}(i,d) + h_{id}^\mu J_{id}^\mu \right) \\ + \sum_{\mu\nu, id, jd'} \mathcal{J}_{id,jd'}^{\mu\nu} J^\mu_{id} (J^\nu_{jd'} - 2\expval{J^\nu_{jd'}})
 \end{multline} 
Note that in the above Hamiltonian we do not refer to a separate ``magnetic sublattice" index, because this is built in to the basis index of the magnetic unit cell. We proceed to summarize the calculations of the well-known random-phase approximation (RPA), which calculates the Fourier transform of the spin-spin correlation $\expval{ \hat{J}^{\alpha}_{id}(t) \hat{J}^{\beta}_{jd'}(0)}$ using the Greens function approach. First, we calculate the non-interacting Greens function matrix $\mathsf{g}(\omega)$
representing the single-ion susceptibility in the mean-field treatment, with matrix elements 
 \begin{small} \begin{equation} \label{eq:NonInteractingGreens}
     g^{\alpha\beta}_{id,jd'}(\omega) = \delta_{ij} \delta_{dd'} \sum_{nm} \frac{\prescript{}{d}{\bra{m}} \hat{J}^{\alpha} \ket{n}_d \prescript{}{d'}{\bra{n}} \hat{J}^{\beta} \ket{m}_{d'}}{\omega+i\epsilon-(\omega_{nd}-\omega_{md'})} (f_{md'} - f_{nd}) 
 \end{equation} \end{small} 
where $ij$ are the indices of the Bravais lattice, $dd'$ are the basis atoms of the magnetic ions, and $H_\mathrm{mf} \ket{n}_d = \omega_{nd} \ket{n}_d$ is the eigenvalue problem for the mean-field Hamiltonian, and $f_{nd}$ is the Boltzmann weight for the state $\ket{n}_d$. The temperature dependence of the Greens function is entirely contained in the Boltzmann weights and the mean-field Hamiltonian. The non-interacting Greens function is a single-ion function and has poles at the energies $(\omega_{nd}-\omega_{md})$ corresponding to excitations of the basis site $d$ from state $m$ to state $n$. Due to the periodicity of the Bravais lattice, here $\hat{J}^\mu_i = \hat{J}^\mu$ are the angular momentum operators with Cartesian component $\mu$, expressed always in the lab frame. We also define the Fourier transform $g_{dd'}^{\alpha\beta}(\vb{q},\omega) = g_{dd'}^{\alpha\beta}(\omega)$ which is $\vb{q}$-independent due to the delta function $\delta_{ij}$ in Eq.\ref{eq:NonInteractingGreens}. 

Within RPA, we obtain a closed form for the expression of the interacting Greens function in the matrix equation 
 \begin{small} \begin{equation}
     \mathsf{G}(\vb{q},\omega) = (I - 2 \mathsf{g}(\omega) \cdot \mathsf{J}(\vb{q}))^{-1} \mathsf{g} (\omega) 
 \end{equation} \end{small} 
where we define the square matrices 
 \begin{gather}
 \mathsf{G}(\vb{q},\omega) = [G_{(\alpha d),(\beta d')}(\vb{q},\omega)] \\ 
 \mathsf{g}(\omega) = [g_{(\alpha d),(\beta d')}(\omega)]\\
\mathsf{J}(\vb{q}) = [\mathcal{J}_{(\alpha d),(\beta d')}(\vb{q})]
 \end{gather}
Here $(\alpha,d)$ are the Cartesian and basis atom indices respectively, resulting in $3|d| \times 3|d|$ matrices, where $|d|=4\times 9 =36$ is the number of sites in the magnetic unit cell. With the interacting Greens function, we write the scattering law via the fluctuation-dissipation theorem as 
 \begin{align} \label{eq:ScatteringLaw}
    S(\vb{q},\omega) = \frac{1}{|d|} \left( \frac{\gamma r_0}{2} \right)^2 g_J^2 |f(q)|^2 \sum_{\alpha\beta} \left( \delta_{\alpha \beta} - \frac{q_\alpha q_\beta}{q^2} \right) \\
    \quad \times -\frac{1}{\pi} \frac{1}{1-e^{-\hbar \omega/k_B T}} \sum_{dd'} \Im G_{dd'}^{\alpha\beta}(\vb{q},\omega+i\epsilon(\omega) ) \nonumber 
 \end{align} 
where $(\gamma r_0/2)^2 \approx 0.0727$ bn relates to the gyromagnetic factor of the neutron; $g_J=8/11$ is the \textit{isotropic} g-factor of Nd$^{3+}$ with total angular momentum $J=9/2$; $|f(q)|^2$ is the magnetic form factor for Nd$^{3+}$; and $\epsilon(\omega)$ is the energy-dependent broadening due to the combined effect of instrument resolution and intrinsic lifetime. 

In the Greens function formalism, the modes follow a Lorentzian distribution due to the imaginary factors $\omega+i\epsilon$ in the denominator. In contrast, the combination of intrinsic lifetime $\eta$ (Lorentzian-distributed) and instrumental-resolution $f(\omega)$ (Gaussian-distributed) is formally a Voigt distribution. Thus, we approximated the Voight distribution as a Lorentzian whose width follows a phenomenological relation \cite{Whiting1968} 
 \begin{equation}
    2\epsilon(\omega) = \eta_{n\vb{q}} + \sqrt{ \eta_{n\vb{q}}^2 + f(\omega)^2 } 
 \end{equation}
In general, the intrinsic lifetime $\eta_{n\vb{q}}$ can be both mode-dependent and wavevector-dependent, but we approximate its effect as a single constant $\eta_{n\vb{q}} \approx \eta$. We find, similar to other metals subject to the Korringa law, that the intrinsic broadening $\eta$ increases with increasing temperature due to the decreased lifetime as a result of electronic scattering \cite{Jensen1991}.

\subsection{Modeling and Fitting} 

In the above, it is assumed that the data are normally distributed, allowing the use of the chi-square to determine the estimates of the parameters. Given the scattering law $S(\vb{q},\omega)$ and a set of parameters $\vb*{\theta}$, we calculate the chi-squared function for the RPA calculation as 
 \begin{equation} 
     \chi_\mathrm{RPA}^2 (\vb*{\theta}) = \sum_{i} \left( \frac{I(\vb{x}_i)/A - S_0(\vb{x}_i,\vb*{\theta})}{ \delta I(\vb{x}_i)/A } \right)^2 
 \end{equation} 
where $\vb{x}_i = (\vb{q}_i,\omega_i)$ are points in $(hhl,\omega)$ space, $I$ and $\delta I$ are respectively the intensity and its associated standard error, and $A\sim 1$ is an amplitude parameter, refined as 1.37(8) due to the uncertainty from the normalization of the (004) peak. For these calculations, we have fit the 11 meV datasets for both 10 K and 0.2 K. 

To obtain a starting set of crystal field parameters, we calculated the scattering law for the non-interacting Greens function in the mean-field approximation, which does not take into account dispersion. We additionally refined two free parameters $\widetilde{h}_1^z$ and $\widetilde{h}_2^z$ to get an estimate of the mean-field energy scale. From these estimates, we could take estimates of exchange couplings that gave consistent mean-field values. 

After obtaining a set of crystal field parameters and estimates of exchange constants, we add further constraints on the chi-square metric to match quantities relating to the measured spin structure, whose excitations lead to the inelastic scattering spectrum. First, we used the method of Luttinger-Tisza to estimate the (incommensurate) magnetic ordering wavevector $\vb{k}_\mathrm{LT}$, and we minimized its deviation from the experimentally determined value $\vb{k}_\mathrm{inc}$. Next, we define the average canting ratio defined by summing the canting ratio over all sites, 
 \begin{equation}
     \bar{\vartheta} \equiv \frac{1}{N_d} \sum_d \frac{\sqrt{\expval{J^x_d}^2 + \expval{J^y_d}^2} }{|\expval{J^z_d}|} 
 \end{equation} 
and likewise minimize the deviation from the observed average of $\bar{\vartheta} \approx \frac{2}{3} \cdot 5\%$. These constraints amount to minimizing the total chi-squared 
 \begin{small} \begin{equation}
     \chi^2(\vb*{\theta}) = \chi^2_{\mathrm{RPA}} + \alpha^2 \left( \frac{|\vb{k}_\mathrm{inc}| - |\vb{k}_\mathrm{LT}(\vb*{\theta})|}{\delta |\vb{k}_\mathrm{inc}|} \right)^2 + \beta^2 \left( \frac{\bar{\vartheta} - \bar{\vartheta}_\mathrm{th}(\vb*{\theta})}{\delta \bar{\vartheta}} \right)^2 
 \end{equation} \end{small} 
The parameters $\alpha$ and $\beta$ for the refinement are arbitrary, but were kept fixed to $\alpha=10^3$ and $\beta = 25\times 10^3$ in order to have substantive effect on the refinement. We proceed to minimize the reduced chi-squared $\chi^2/(N-N_0)$ where $N=182370$ pixels and $N_0 = 27$ parameters. We used the Nelder-Mead (heuristic) minimizer to refine each parameter freely. While Nelder-Mead does not guarantee a global minimum (and can fall into saddle points), we found it converged better than the Levenberg-Marquardt least-squares optimizer due to the possibility of local extrema. 

To estimate uncertainties in the parameters, we performed a Monte Carlo calculation. This was aided by the principal component analysis of an initial set of variations to the exchange constants \cite{Parente2013, Scheie2022}. In our Monte Carlo algorithm, we accept the step on the condition that $z<e^{-\chi^2/t}$ for random number $z\in [0,1]$ and Monte Carlo temperature $t=0.25$, and that $\chi_\mathrm{min}^2$ does not exceed a given threshold (discussed below). Despite the large $\alpha,\beta$ coefficients, the contributions of these costs on the reduced chi-square were on average less than 1\% of the RPA cost; therefore, the parameter sweep was essentially done in the subspace of constant wavevector and canting angle. 

The sum of $\nu$ (degrees of freedom) Gaussian random variables forms a chi-squared distribution $\chi_\nu^2$ with mean $\nu$, and one standard deviation uncertainty in the parameters corresponds to a change of $\Delta \chi_\nu^2 = 1$ \cite{Barlow2004, Martin2012}. Therefore, the reduced chi-squared statistic $\chi_\mathrm{red}^2 = (1/\nu) \chi_\nu^2$ is expected to have mean of 1 and one standard deviation uncertainty in parameters has $\Delta \chi_\mathrm{red}^2 = 1/\nu$. Therefore, parameters ought to be swept until the condition $\chi_\mathrm{max}^2 = \chi_\mathrm{red}^2 (1+1/\nu)$ is reached. However, in the case of the RPA calculation, the large number of fitted points $N=182370$ makes $1/\nu = 1/(N-N_0)$ exceptionally small. As a compromise, we conditioned the Monte Carlo simulation to keep parameters below $\Delta \chi_\mathrm{red}^2 = 1$ and reported the mean and standard deviation of the parameters in the Monte Carlo sweeps as the best fit values.

\section{Contributions}
The analysis of the experimental data was performed by C.J.L. under the supervision of C.L.B. The crystals were grown by H-Y.Y. and X.Y. under the supervision of F.T. The inelastic neutron scattering experiments were performed by C.J.L. and J.G. with assistance from A.P. (CNCS) and J.A.R. (NIST). The diffraction measurements were performed by C.J.L. with assistance from Y.H. and H.C. (DEMAND). S.B. provided the DFT and spiralization tensor analysis, and P.N. contributed the field theory analysis. All authors provided feedback and shaped the research, analysis and manuscript.

\section{Acknowledgments} 
We would like to thank Santu Baidya for providing DFT calculations of the Weyl nodes; Juba Bouaziz for providing the DFT magnetic interactions and fruitful discussion; Julie Staunton and Christopher Patrick for providing the DFT calculations of crystal field parameters and valuable discussions; and Andrew Boothroyd and Timothy Reeder for valuable discussions. This work was supported as part of the Institute for Quantum Matter, an Energy Frontier Research Center funded by the U.S. Department of Energy, Office of Science, Basic Energy Sciences under Award No. DE-SC0019331. 
C.B. was supported by the Gordon and Betty Moore Foundation EPIQS program under GBMF9456. 
The work at Boston College was funded by the U.S. Department of Energy, Office of Basic Energy Sciences, Division of Physical Behavior of Materials under award number DE-SC0023124.
Computational analysis was carried out at the Advanced Research Computing at Hopkins (ARCH) core facility (rockfish.jhu.edu), which is supported by the National Science Foundation (NSF) grant number OAC1920103. 
A portion of this research used resources at the Spallation Neutron Source, a DOE Office of Science User Facility operated by the Oak Ridge National Laboratory.
A portion of this research used resources at the High Flux Isotope Reactor, a DOE Office of Science User Facility operated by the Oak Ridge National Laboratory.
Access to MACS was provided by the Center for High Resolution Neutron Scattering, a partnership between the National Institute of Standards and Technology and the National Science Foundation under Agreement No. DMR-2010792. The identification of any commercial product or trade name does not imply endorsement or recommendation by the National Institute of Standards and Technology.

\section{Data Availability}
The data that support the findings of this work are available from the corresponding author upon request. 


\counterwithin{figure}{section}
\renewcommand\thefigure{S\arabic{figure}} 

\section{Supplementary Information (SI)}
\subsection{Magnetic structure analysis} 
We use the convention that the real-space magnetic structure $\vb{m}(\vb{r}_{id})$ is expressed as \cite{RODRIGUEZCARVAJAL1993}
 \begin{equation} \label{eq:FourierSpinComponents}
     \vb{m}(\vb{r}_{id}) = \sum_{\{\vb{k}\} } \vb{T}_{d}(\vb{k}) e^{-i\vb{k}\cdot \vb{R}_{id}} = \sum_{\{\vb{k}\}} \vb{S}_{d}(\vb{k}) e^{-i\vb{k}\cdot \vb{R}_{i}},
 \end{equation}
where $\vb{T}_d(\vb{k})$ and $\vb{S}_d(\vb{k})$ are Fourier amplitudes, $\{\vb{k}\}$ are the set of wavevectors making up the star of the magnetic propagation vector (nominally $\pm \vb{k}_\mathrm{mag}$), $\vb{R}_{id} = \vb{R}_i + \vb{r}_d$, such that $\vb{R}_i$ is a Bravais lattice vector and $\vb{r}_d$ is the atomic basis in the conventional tetragonal unit cell. While $\vb{S}_d(\vb{q})$ is commonly referred to in the literature and analysis software like Fullprof and SARAh Refine, we find it convenient to refer to $\vb{T}_d(\vb{q})$ due to the form of the Fourier transform which is used in the Luttinger-Tisza model. 

An important result of the symmetry relations is how the structures of the ordered moment are related between sites. The local coordinates of the two Nd sites in the primitive unit cell are related by exchanging $x$ and $y$ due to the non-symmorphic mirror symmetry operation $g_{8} = \sigma_{db} ' = \left\{\sigma_{db} | ~ y-\frac{1}{2}, x-1, z+\frac{3}{4} \right\}$ in the little group for magnetic structure with propagation vector $\vb{k}_\mathrm{com}=(2/3, 2/3, 0)$. We write $\vb{T}_d = \sum_j c_j \vb*{\psi}_{jd}$ where $c_j$ are generally complex coefficients and $\vb*{\psi}_j$ are the basis vectors of the irreducible representation for the particular spin structure. The symmetry and space group of the ordered state can be written in the magnetic space group approach. Here, the symmetry operations consistent with the spin structure are the identity and $\bar{g}_8$ where $\bar{x}$ denotes time-reversal conjugation. The operations are consistent with the space group $Cc'$ (\#9.39) whose point group is $m'$ (\#4.3.11). Note that this point group is compatible with ferromagnetism, as we expect. Here the Fourier components will satisfy $\vb{T}_1 = (T_x, T_y, T_z)$ and $\vb{T}_2 = (T_y, T_x, T_z)$ where $T_x,T_y,T_z$ are in general complex numbers. From the diffraction data, we write these Fourier components in the convenient notation 
 \begin{small} \begin{equation}
     \vb{T}_1(\vb{k}_\mathrm{com}) = \mqty( i\delta \cos \varphi/\sqrt{3} \\ i\delta \sin \varphi/\sqrt{3} \\ \eta/2 ), \quad \vb{T}_2(\vb{k}_\mathrm{com}) = \mqty( i\delta \sin \varphi/\sqrt{3} \\ i\delta \cos \varphi/\sqrt{3} \\ \eta/2 ) \end{equation} \end{small} 
The Fourier component for the ferromagnetic component is simply $\vb{T}_d(\vb{k}_0=\vb{0}) = (0,0,-\xi)$. Therefore in this convention, the moment at $(0,0,0)$ is $(0,0,\eta-\zeta)$, at $(1,0,0)$ it is $(-\delta \cos \varphi, -\delta \sin \varphi, -\zeta-\eta/2)$ and at $(1/2,0,1/4)$ is $(\delta \sin \varphi, \delta \cos \varphi, -\zeta-\eta/2)$. 

\begin{figure}
    \centering 
    \includegraphics[width=\linewidth]{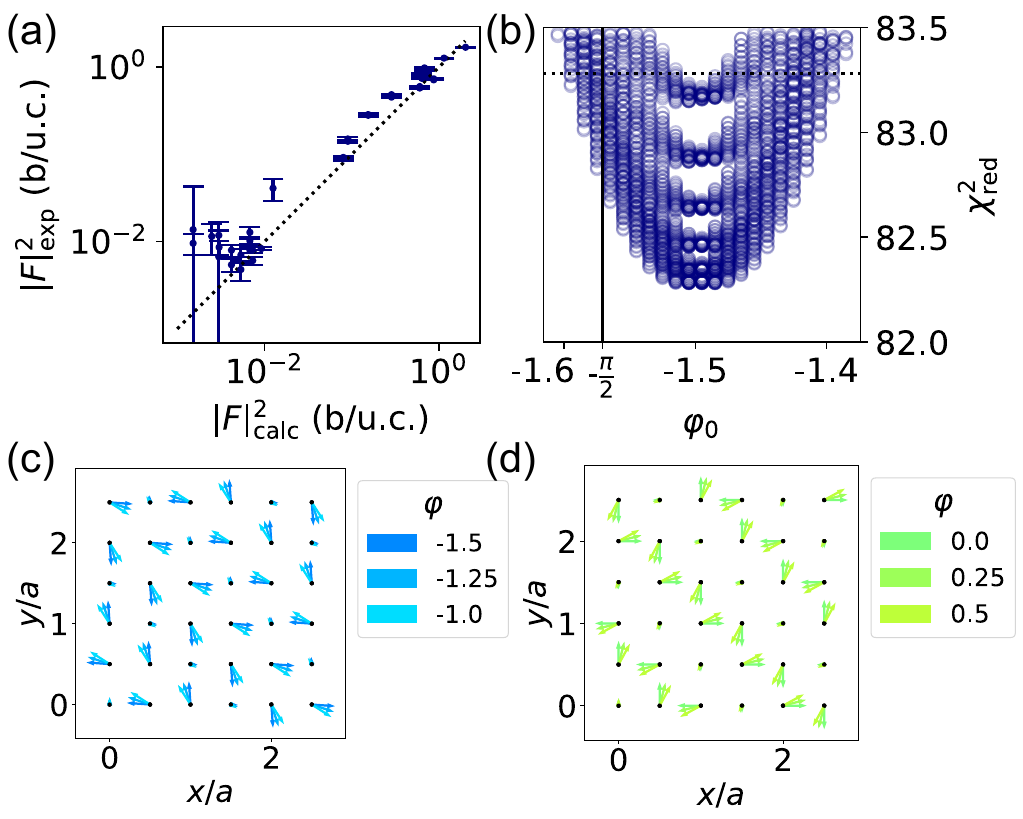} 
    \caption{ (a) Theoretical and experimental structure factors. (b) Minimum chi-squared by sweeping the parameters $\delta,\eta, \varphi$ away from the optimized values, revealing the shallow minimum near $\varphi=-1.5(1)$. (c,d) Example spin structures for varying $\varphi$, showing the in-plane component of spinss aligned mainly perpendicular or parallel to the wavevector $(\tfrac{2}{3} \tfrac{2}{3} 0)$, respectively. } 
    \label{fig:SI_Diffraction} 
\end{figure}

The neutron diffraction data are fit by calculating 
 \begin{equation}
     S(\vb{Q}) = (\gamma r_0/2)^2 |f(Q)|^2 \sum_{\alpha \beta} 
 (\delta_{\alpha\beta} - \hat{q}_\alpha \hat{q}_\beta) (F^\alpha(\vb{Q}))^* F^\beta(\vb{Q})
 \end{equation}
where $|f(Q)|^2$ is the squared form factor of Nd$^{3+}$, and the magnetic vector $\vb{F}(\vb{Q})$ is defined as 
 \begin{equation}
     F^\mu(\vb{Q}) = F^\mu(\vb{H}+\vb{k}) = \sum_d T_{d}^\mu(\vb{k}) e^{+i\vb{H}\cdot \vb{r}_d} 
 \end{equation} 
with the sum over the basis atoms of the unit cell, and we explicitly separate the reciprocal lattice vector $\vb{H}$ from the magnetic wavevector $\vb{k}$. Fig.~\ref{fig:SI_Diffraction} shows the results of the magnetic structure refinement highlighting the agreement of the weak peaks, which are most sensitive to the canting angles. 

\subsection{Crystal field, DFT, and parameter estimations} 
The orthorhombic crystal field symmetry manifests in low-symmetry wavefunctions for the ground state and excited states. The general form of the state $\ket{n}$ is $\ket{n} = \sum_m c(n,m) \ket{m}$ where $\ket{m} \equiv \ket{J,m}$ are the states of definite projection of angular momentum and the sum is over all $-J\leq m \leq J$. In general $c(n,m)\neq 0$, so that there are contributions from all projections of angular momentum. Indeed, all crystal field states share this general form as they all lie in the same irreducible representation $\Gamma_5$ formed in the double group of $C_{2v}$ \cite{Koster1963}. An Ising-like ground state $\ket{\pm}$ would correspond to large weights $c(\pm, \pm 9/2)$, but in general the nonzero weights on the $m\neq \pm 9/2$ components will reduce the ground-state moment from the maximal value $g_J \mu_B J$. Meanwhile, the mean fields or Zeeman fields will split the doublet and put more weight on the $c(0, \pm 9/2)$ component leading to larger magnitude of moment. The sign of the mean field will select the (opposite) sign of the component that has larger contribution, which results in the antiparallel aligment of the spin expectation value with the mean field.

In previous works \cite{Suzuki2019}, the local anisotropy of the sites was expressed using the $g$-tensors $g_1 = \mathrm{diag}(g_x, g_y, g_z)$ and $g_2 = \mathrm{diag}(g_y, g_x, g_z)$. In our work, we refer to the isotropic g-factor $g_J$. The distinction arises from our use of the full crystal field scheme where the moment is $\vb{m} = g_J \mu_B \vb{J}$, whereas the $g$-tensor formalism works in the projection of the Zeeman energy $(-\vb{B}\cdot \vb{m})$ to the ground state doublet $\ket{\pm}$ represented in the Pauli pseudospin basis $(-\vb{B} \cdot g_1 \cdot \vb*{\sigma})$. Our model has the benefit of being able to directly calculate the $g$-tensor from the result of our crystal field fits: 
\begin{multline} 
    g_1 = g_J \mqty( \Re \bra{+} J_x \ket{-} & \Im \bra{+} J_x \ket{-} & \bra{+} J_x \ket{+} \\ \Im \bra{+} J_x \ket{-} & \Im \bra{+} J_y \ket{-} & \bra{+} J_y \ket{+} \\ \bra{+} J_x \ket{+} & \bra{+} J_y \ket{+} & |\bra{\pm} J_z \ket{\pm}| ) \\
    = \mqty( g_{xx} & g_{xy} & g_{xz} \\ g_{xy} & g_{yy} & g_{yz} \\ g_{xz} & g_{zy} & g_{zz} ) = g_J \mqty( 0.250 & 0 & 0 \\ 0 & 1.149 & 0 \\ 0 & 0 & 3.407) 
\end{multline}
Under the $4^\pm$ nonsymmorphic symmetry, the $g$-tensor will transform as $g_d = R_d^{-1} g_1 R_d$ where $R_d$ is the classical $\phi_d$ rotation matrix about the $z$-axis. This equivalently can be explained due to the quantum mechanical rotation of the eigenfunctions $\ket{\pm}_d$. 

\begin{figure}
    \centering
    \includegraphics[width=\linewidth]{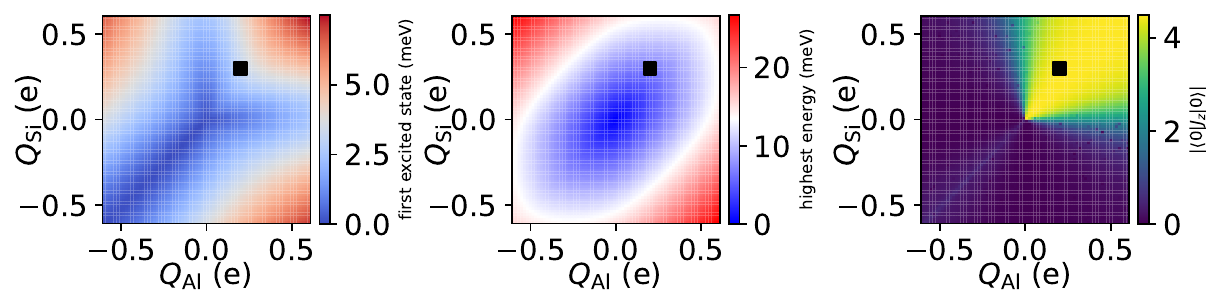}
    \caption{ Screened point charge model from sweeping freely the two indendent charge parameters $Q_{\mathrm{Si}}$ and $Q_{\mathrm{Al}}$ to closely match the first excitated states, highest energy level, and the $z$-projection of ground state wavefunction.}
    \label{fig:SI_ScreenChargeModelSpace}
\end{figure}

The initial crystal field parameters for refinements were estimated in two ways, each of which produced similar preliminary crystal field refinements. First, we used the crystal field coefficients found in \cite{Bouaziz2024} as an initial parameter set for the refinement. We note that the coefficients with large $n$ tend to be inaccurate due to the large angular momentum, and as a result will typically under-estimate the bandwidth of the crystal field spectrum. The coefficients were calculated as $B_{nm} = \alpha_{nm} \theta_n [A_{nm} \expval{r^n}]$, where $\alpha_{nm}$ are coefficients found in converting from the Wybourne coefficients. The reduction of the bandwidth of the states required a scaling to reach reasonable fitting, for example a single scaling parameter $\alpha H_{\mathrm{DFT}}$, where $\alpha=1.5-2.0$ gave reasonable starting parameters to perform fits. 

\begin{figure}
    \centering
    \includegraphics[width=0.75\linewidth]{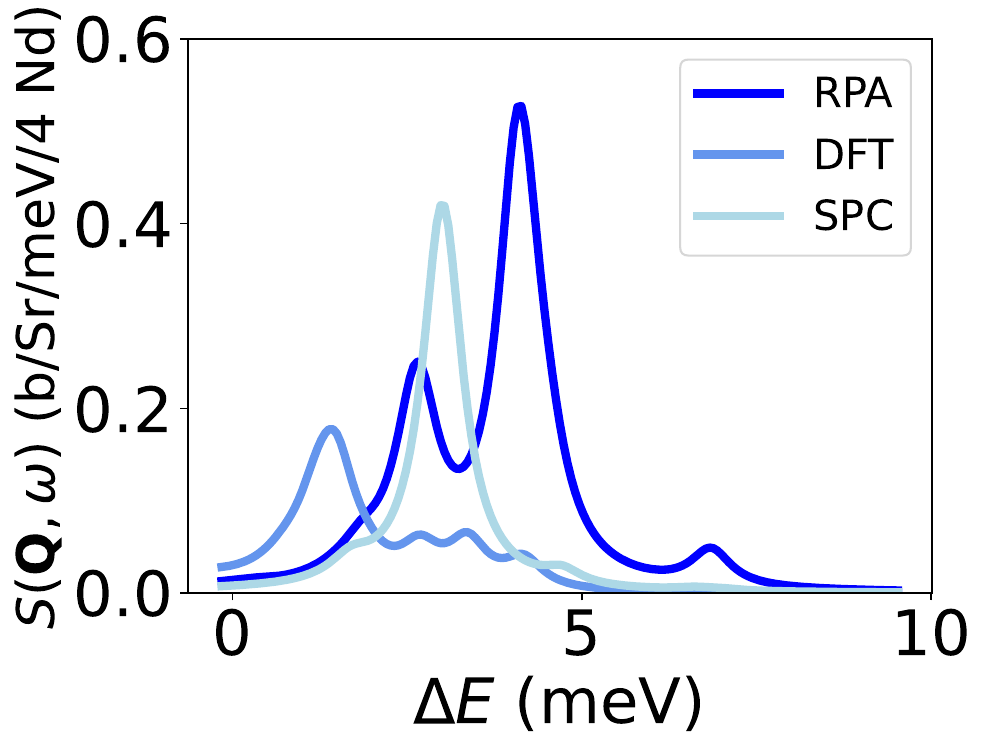}
    \caption{ Comparison of the calculated paramagnetic (10 K) spectrum near $\vb{Q}=(\tfrac{2}{3} \tfrac{2}{3} 6)$ for the experimental fit (EXP), DFT, and SPC models in absolute units.}
    \label{fig:SI_RPA_CEF_comparison}
\end{figure}

Next, we used a screened point-charge calculation by taking the local coordination of aluminum and silicon ions above and below the local Nd site, forming the $C_{2v}$ symmetric cage. We take free charges for $q_{\mathrm{Al}}$ and $q_{\mathrm{Si}}$ as free parameters to calculate the Stevens operator coefficients \cite{Hutchings1964} 
 \begin{equation}
     B_{nm} = (-ke^2) c_{nm} \expval{r^n} \theta_n \sum_{j} \frac{4\pi}{2n+1} \frac{q_j}{r_j^{n+1}} Z_{nm}^c(\theta_j, \phi_j)
 \end{equation} 
with $q_j$ in units of $e$, where $k$ is the Coulomb constant, $(-e)$ is the charge of the electron, $c_{nm}$ are coefficients that relate the tesseral $Z_{nm}^c(\theta,\phi)$ and spherical $Y_{nm}(\theta,\phi)$ harmonics, $\theta_n$ are $J$-dependent Stevens coefficients $\{\alpha_J, \beta_J, \gamma_J\}$ for $n=2,4,6$ respectively, and $(r_j,\theta_j,\phi_j)$ are the spherical coordinates of the ligands at position $\vb{r}_j$. The $q_j$ represent the effective charges perhaps due to screening from itinerant electrons. As shown in Fig.~\ref{fig:SI_ScreenChargeModelSpace}, we vary the two charge parameters to explore the ground state doublets (expected $\ket{\pm 9/2}$ states), the energy of the first excited state doublet, and the energy of the maximum excited state. These inform regions of this reduced space for which to start the refinement.

\subsection{Luttinger-Tisza Method} 
Any exchange parameters extracted from excitations in INS data must be consistent with the ground state spin structure that yield the excitations in the first place. To do this, we use the method of Luttinger-Tisza \cite{Schmidt2022}, which is a classical spin model that arises in relaxing the fixed spin-length constraint that occurs in the minimization problem 
 \begin{small} \begin{equation}
     E = E_{\mathrm{CEF}} + \sum_{id,jd'} \mathcal{J}_{idjd'} J^\mu_{id} J^\nu_{jd'} - \lambda \left( \sum_{id} |\vb{J}_{id}|^2 - NJ^2 \right) 
 \end{equation} \end{small} 
The Hamiltonian is expressed in the paramagnetic state where we do not assume the moments yet have a finite expectation value; the atomic basis $d,d'\in \{d_0\}$ is the chemical basis of the unit cell with 4 dependent sites. The CEF energy, to second order in the angular momentum operators, is
 \begin{small} \begin{equation}
   E_{\mathrm{CEF}} = \sum_{id} (-K_{20}) (J_{id}^z)^2 + (-1)^d K_{22} [(J_{id}^x)^2 - (J_{id}^y)^2 ] 
 \end{equation} \end{small} 
where $-K_{20} = 3B_{20}$ is assumed to be negative to give easy-axis anisotropy, $K_{22} = B_{22}$ can have any sign, and the factor $(-1)^d$ arises from the rotation operation in Eq.(\ref{eq:QM_Rotation_Hamiltonian}). 
Upon taking the Fourier transform,  
 \begin{equation} 
     J_{id}^\mu = \sum_{\vb{q}} J_{d}^\mu (\vb{q}) e^{i\vb{q}\cdot \vb{r}_{id}}
 \end{equation} 
and minimizing the energy with respect to variations $\delta E/\delta (J_{kd''}^\alpha)$, we obtain the eigenvector problem 
 \begin{equation}
      [{\mathcal{J}}^{\mu\nu}_{dd'}(\vb{q}) + \mathcal{K}_{dd'}^{\mu\nu}] J^{\nu}_{d'}(\vb{q}) = \lambda(\vb{q}) J^{\mu}_{d}(\vb{q}) 
 \end{equation}
Here $\mathcal{K}_{dd'}^{\mu\nu} = \delta_{dd'} \delta^{\mu\nu} [\delta^{\mu z} (-K_{20}) + (-1)^d K_{22} (\delta^{\mu x} - \delta^{\mu y})]$ reflect the single-ion energies. The exchange Hamiltonian is Hermitian, including the antisymmetric terms, leading to an eigenvalue problem for a fixed wavevector that yields a set of ordered real energies $\{ \lambda_p(\vb{q}) \}$. To minimize the system energy, we seek to find the wavevector that minimizes the minimum eigenvalue, namely $\min_{\vb{q}} \{\lambda_0(\vb{q}) \}$. Once a wavevector is chosen, denoted $\vb{k}_{\mathrm{mag}}$, the eigenvector $\{ J_{d}^\mu (\vb{k}_{\mathrm{mag}}) \}$ describes the Fourier transform of the angular momentum vectors for the different basis atoms. This is in fact the estimate of the vectors $\vb{T}_{d}(\vb{q})$ in Eq.(\ref{eq:FourierSpinComponents}). In summary, the estimates of the crystal field and exchange parameters directly predict the magnetic wavevector and spin structure. 

Often, for incommensurate spin systems, the Luttinger-Tisza method leads generically to spin spirals that can have unequal magnitude on each spin site. Such a solution will not correspond to the true $T=0$ ground state for Kramers ion systems as it does not satisfy the strong constraint. This is the case in NdAlSi, where the predicted spin structure (neglecting spin canting) is the basis vector $\Gamma_1(\psi_3)$ with $m_1^z = +m$ and $m_2^z = -m/2$. We thus treat the predicted wavevector and spin structure from Luttinger-Tisza as that arising near $T_\mathrm{incom}$. For $T<T_\mathrm{com}$, we use the Fourier components as an initial estimate to compute the self-consistent mean-field spin structure in a commensurate unit cell, whereby the spin length is controlled by the crystal field Hamiltonian. Simultaneously, there sometimes exist frustrated exchange parameters that make the Luttinger-Tisza method unstable, or produce degenerate or even spin-liquid states \cite{Halloran2023}. However, taking the Luttinger-Tisza result into the mean-field model yields a self-consistent and convergent solution to the ground state order. 

While the above eigendecomposition problem may seem intractable given its size $3d_0 \times 3d_0$, it is informative to get an analytic estimate of the minimum wavevector. Without the DM interaction, and assuming an XXZ Hamiltonian, the minimization of the interaction matrix $\mathcal{J}_{dd'}^{\mu\nu}$ can be done by considering only the block diagonal $\mathcal{J}_{dd'}^{zz}$, since the eigenvalues of this block will be minimized due to the inclusion of the single-ion term $-K_{20} J_z^2$. This block matrix has the form 
 \begin{equation}
     \mathcal{J}_{dd'}^{zz}(\vb{q}) = \mqty( c_{\vb{q}} - K_{20} & a_{\vb{q}} & d_{\vb{q}} & b_{\vb{q}} \\ a^*_{\vb{q}} & c_{\vb{q}} - K_{20} & b^*_{\vb{q}} & d_{\vb{q}} \\ d_{\vb{q}} & b_{\vb{q}} & c_{\vb{q}} - K_{20} & a_{\vb{q}} \\ b_{\vb{q}}^* & d_{\vb{q}} & a_{\vb{q}}^* & c_{\vb{q}} - K_{20} )
 \end{equation}
where 
 \begin{multline}
     c_{\vb{q}} - K_{20} = 3B_{20} + 2\mathcal{J}_2^z [\cos(2\pi h) + \cos(2\pi k)] \\
     + 2\mathcal{J}_4^z [ \cos 2\pi(h+k) + \cos 2\pi(h-k) ] + \cdots 
 \end{multline}
 \begin{multline}
     a_{\vb{q}} = 2e^{-i\pi l/2} [\mathcal{J}_1^z \cos \pi k + 2\mathcal{J}_3^z \cos 2\pi h \cos \pi k \\
     + \mathcal{J}_5^z \cos 3\pi k + 2\mathcal{J}_7^z \cos 3\pi h \cos 2\pi k + \cdots ] 
 \end{multline}
 \begin{multline}
     b_{\vb{q}} = 2e^{+i\pi l/2} [\mathcal{J}_1^z \cos \pi h + 2\mathcal{J}_3^z \cos 2\pi k \cos \pi h \\
     + \mathcal{J}_5^z \cos 3\pi h + 2\mathcal{J}_7^z \cos 3\pi k \cos 2\pi h + \cdots ] 
 \end{multline}
 \begin{multline}
     d_{\vb{q}} = 8\mathcal{J}_6^z \cos \pi h \cos \pi k \cos \pi l \\ + 8\mathcal{J}_{11}^z [\cos \pi h \cos 3 \pi k + \cos 3\pi h \cos \pi k] \cos \pi l + \cdots \label{eq:DQ}
 \end{multline}
The four eigenvalues of this matrix are $\lambda (\vb{q}) = (c_{\vb{q}} - K) \pm d_{\vb{q}} - |a_{\vb{q}} \pm b_{\vb{q}}|$. Along $\vb{q}=(hh0)$ the minimum eigenvalue is (written up to nine nearest neighbors) 
\begin{small} 
\begin{multline} \label{eq:MinimumWavevector}
    \lambda_0(hh0) = (3B_{20} - 2|\mathcal{J}_4| - 4|\mathcal{J}_6|) \\
    + \bigg[4\left( \mathcal{J}_2^z - |\mathcal{J}_6^z| \right) \cos 2\pi h + 2(\mathcal{J}_4^z + \mathcal{J}_8^z) \cos 4\pi h + \cdots \bigg] \\
    - 4 \bigg| \left(\mathcal{J}_1^z + \mathcal{J}_3^z + \mathcal{J}_7^z \right) \cos\pi h + \left(\mathcal{J}_3^z + \mathcal{J}_5^z + \mathcal{J}_9^z \right) \cos3\pi h \\ 
    + (\mathcal{J}_7^z + \mathcal{J}_9^z) \cos 5\pi h + \cdots \bigg| 
\end{multline}
\end{small} 

Of course, with a large set of exchange parameters, it is not guaranteed that $(hh0)$ will be the optimal wavevector compared to $(h00)$, for example. The minimization routine of the full matrix $\mathcal{J}_{dd'}^{\mu\nu}(\vb{k}_\mathrm{mag} )$ for the magnetic wavevector $\vb{k}_\mathrm{mag}$ is ultimately done numerically. However, the above exercise illustrates broadly speaking how the competition of the various exchange constants will generally lead to incommensurability. Furthermore, it can give an indication of when to expect proximity to certain minima. Note $\vb{k}_\mathrm{mag} \approx (\tfrac{2}{3} \tfrac{2}{3} 0)$ is equivalent to $\vb{k}_\mathrm{mag} \equiv \vb{G}_3 - \vb{k}_\mathrm{mag}'$ where $\vb{k}_\mathrm{mag}' \approx (\tfrac{1}{3} \tfrac{1}{3} 0)$ lies in the first Brillouin zone and $\vb{G}_3 = (110)$ is a reciprocal lattice vector of the primitive lattice. These correspond to the extremum of the $\cos 3\pi h$ term when $h=1/3$. Up to nine nearest neighbors, this harmonic only appears in $|a_{\vb{q}} + b_{\vb{q}}|$. Focusing on this term, suppose $(\mathcal{J}_7^z + \mathcal{J}_9^z) \approx 0$ for illustration. Then for instance if $A=(\mathcal{J}_1^z + \mathcal{J}_3^z + \mathcal{J}_7^z)$ and $B=(\mathcal{J}_3^z + \mathcal{J}_5^z + \mathcal{J}_9^z)$ have the same sign, $h=0$ will be the minimum value of this expansion. But if $A$ and $B$ have opposite signs, then $h \sim 1/3$ is the minimum value of this expansion when $0 < -A/3B < 3$. This term will then compete with other harmonics when the full functional form is taken into account, but this discussion serves to demonstrate that sign changes in the exchange constants are an important ingredient in obtaining a magnetic propagation vector close to $(\tfrac{2}{3} \tfrac{2}{3} 0)$. 


\subsection{ Exciton spectrum and dispersion }
To illustrate the role of the exchange interaction when the exchange is comparable but smaller than the crystal field energy level scheme, we can consider the case of an isotropic spin system with one site per unit cell and isotropic interactions. Here the RPA gives a Greens function that is a scalar, 
 \begin{equation}
     G(q,\omega) = \frac{g_0(\omega)}{1 - 2 g_0(\omega) \mathcal{J}(q)} 
 \end{equation}
where we have now scalar division instead of matrix inversion. At $T=0$, we have $g_0(\omega) = \sum_{n} |M_n|^2/(\omega +i\epsilon - \omega_{n})$ where $|M_n|^2$ is a matrix element. Then the energy modes of the spin spectrum are the poles of the Greens function $G(q,\omega)$, which occur when $1-2g_0(\omega) \mathcal{J}(q) = 0$, or $\omega_n(q) \approx \omega_n + 2|M_{n}|^{2} \mathcal{J}(q)$ if the separation of bands are much larger than the interaction energies. Similar to the tight-binding model, the frequency of the excitations should thus roughly follow the functional form contained in $\mathcal{J}(\vb{q})$, with bandwidth bound by the matrix element. Of course, this picture is lost when looking at the full matrix formulation where bands arising from the different basis atoms and Cartesian components can mix. 

Indeed, the dispersion can be found by diagonalizing the Bogoliobuv Hamiltonian written to quadratic order of the operators that create and destroy mean-field states $\ket{n}$. This is akin to a tight-binding calculation performed in a basis of degenerate states. The energy eigenvalues are found after diagonalizing the symplectic matrix \cite{Buyers1971, Colpa1978, Gaudet2023} 
 \begin{equation} 
     H = \mqty( \mathsf{A}(\vb{q}) & \mathsf{B}(\vb{q}) \\ -\mathsf{B}^*(-\vb{q}) & -\mathsf{A}^*(-\vb{q}) )
 \end{equation}
whose matrix elements are   
 \begin{small}  \begin{gather}
     A_{pp',dd'}(\vb{q}) = \omega_p \delta_{pp'}\delta_{dd'} + 2\sum_{\mu\nu} {\mathcal{J}}_{dd'}^{\mu \nu}(\vb{q}) \prescript{}{d}{\bra{p}} J^{\mu} \ket{0}_d \prescript{}{d'}{\bra{0}} J^{\nu} \ket{p'}_{d'} \\ 
     B_{pp',dd'}(\vb{q}) = 2\sum_{\mu\nu} {\mathcal{J}}_{dd'}^{\mu \nu}(\vb{q}) \prescript{}{d}{\bra{p}} J^{\mu} \ket{0}_d \prescript{}{d'}{\bra{p'}} J^{\nu} \ket{0}_{d'} 
 \end{gather}  \end{small} 

From the above equations, we can anticipate that the largest matrix elements (which couple between sites and different excited states) will preferentially select exchange terms $J^{\mu\nu}_{dd'}(\vb{q})$ to dominate the dispersion. This detail is quite useful in distinguishing models like the Heisenberg, XXZ, XY, and Ising spin models, as the bandwidths will be affected by the components of spins that are coupled. For instance, an Ising model would have less-dispersive bands than a Heisenberg model. 

Since the dispersion of the modes will go roughly as $\mathcal{J}(q)$, we can use this to predict the expected exchange constants necessary to account for the data. For instance, the presence of modes following approximately $\omega(00l) \sim \cos \pi l = \cos (\vb{q}\cdot \vb{R}_6)$ along the $(00l)$ direction implies a coupling of at least sixth-nearest neighbor, $\mathcal{J}_6$, as seen in Eq.~(\ref{eq:DQ}).

\subsection{Spiralization tensor}
We note that the anisotropic DM interaction can in principle lead to stable antiskyrmion phases as discussed in \cite{Hoffmann2017}. As an illustration, using the spin structure of our system, consider the structure written in a continuum limit for a single plane of the square lattice of Nd ions. The spin structure can be written  
 \begin{equation}
     \vb{m}(\vb{r}_{id}) = \sum_{\{\vb{k}\}} \vb{T}_d(\vb{k}) e^{-i\vb{k}\cdot \vb{r}_{id}} 
 \end{equation}
For a $2q$ structure the sum is over the vectors $\vb{k}_\parallel$ and $\vb{k}_\perp$ which are related by paramagnetic space group symmetry
 \begin{align}
   \vb{m}(\vb{r}_{id}) &= [ \vb{T}_d(\vb{k}_\parallel) e^{-i\vb{k}_\parallel \cdot \vb{r}_{id}} + \vb{T}_d^\dagger(\vb{k}_\parallel) e^{+i\vb{k}_\parallel \cdot \vb{r}_{id}} ] \\
   &\quad + [e^{-i\alpha} \vb{T}_d(\vb{k}_\perp) e^{-i\vb{k}_\perp \cdot \vb{r}_{id}} + e^{+i\alpha} \vb{T}_d^\dagger(\vb{k}_\perp) e^{+i\vb{k}_\perp \cdot \vb{r}_{id}} ] \nonumber 
 \end{align} 

In CeAlGe \cite{Puphal2020}, meron-antimeron spin texture was reported and evidenced by topological Hall effect in applied magnetic fields. The latter arises due to an imbalance of the population of merons and antimerons. We can similarly simulate the Fourier transform of the spin structure for $(k00)$ order using Luttinger-Tisza. To give a faithful representation of the magnetic order and exchange interactions, we take a toy model consisting of an easy-plane structure with $K_{20} > 0$ and $K_{22}<0$; symmetric exchange interaction $\mathcal{J}^x_1<0$ and antisymmetric exchange $|D_2| < |\mathcal{J}^x_1|$ whose spiralization tensor has $\det \mathcal{D}<0$, while all other interactions are set to zero. These will stabilize either $\vb{k}=\vb{0}$ or incommensurate $\vb{k}=(k00)$ depending on the relative size of $|D_2/\mathcal{J}^x_1|$. In the incommensurate $2q$ structure, we find 
 \begin{equation} 
     \vb{m}_1(\vb{r}) = \mqty( \eta_1 \sin kx \\ \eta_2 \sin ky \\ \delta_1 \cos kx + \delta_2 \cos ky) 
 \end{equation} 
This form necessarily depends on the sign of $\det \mathcal{D}$. When $\det \mathcal{D}<0$ we find $\eta_1 \eta_2 >0$ and $\delta_1 \delta_2 < 0$, and remarkably this has identical form to that refined in \cite{Puphal2020}. When $\det \mathcal{D}>0$, we get a spin structure with $\eta_1 \eta_2 > 0$ and $\delta_1 \delta_2 > 0$, which is not consistent with the refined data. The change in sign of $\delta_1 \delta_2$ manifests similarly to the change in orientation of the in-plane canting of NdAlSi. We note that our Luttinger-Tisza method does not differentiate in the energetics of the real-space skyrmion versus antiskyrmion structures. However, our method has predictive power for demonstrating which spin textures may develop from a certain spiralization tensor, and how $\det \mathcal{D} < 0$ interactions can stabilize the spin structures found in this family of Weyl semimetals. 

\begin{figure}
    \centering
    \includegraphics[width=\linewidth]{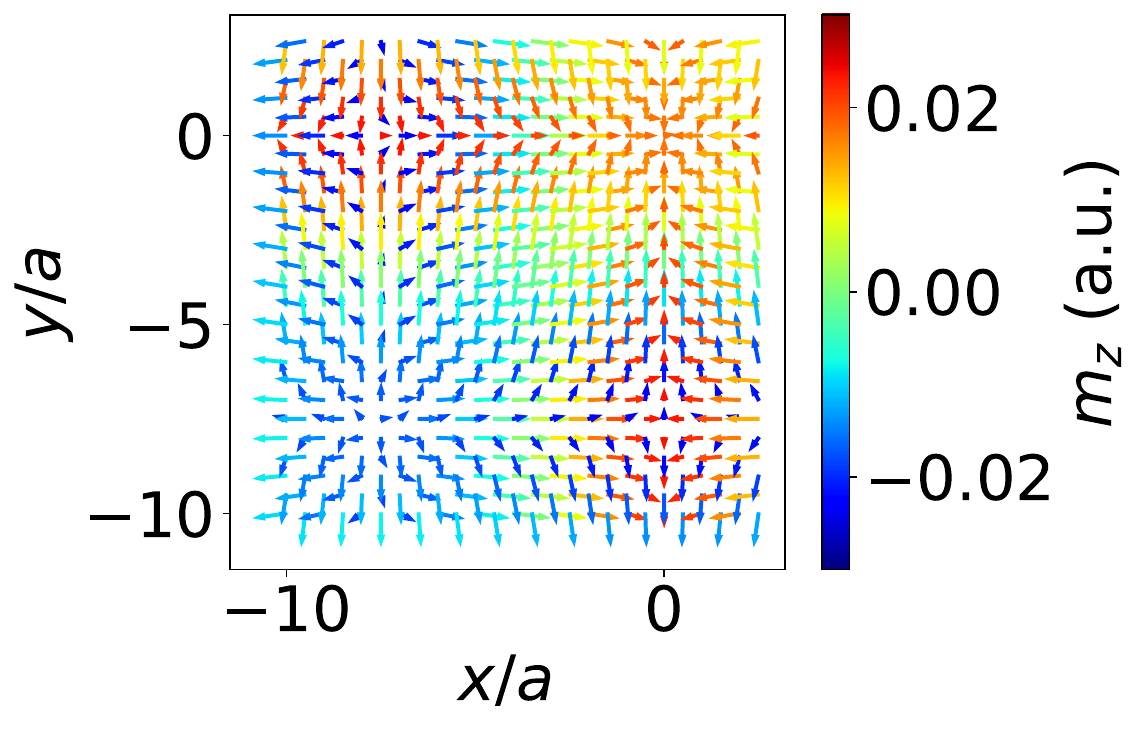}
    \caption{Simulation of the $2q$ spin structure for CeAlGe. Here we took $\mathcal{J}_1^x = -3$ $\mu$eV, $|D_2| = 2.25$ $\mu$eV, $K_{20} = 0.132$ meV and $K_{22} = -0.04$ meV. This given parameter set stabilizes an incommensurate $(k00)$ structure with $k\approx 0.066$ and a negative-determinant spiralization tensor. With these, we find the topological meron-antimeron structure; we plot the spatial range to show consistency with \cite{Puphal2020}.} 
    \label{fig:SI_CeAlGe_Simulation} 
\end{figure}

\subsection{Chi-squared and correlations} 

\begin{figure*}[ht!]
    \centering
    \includegraphics[width=1.0\linewidth]{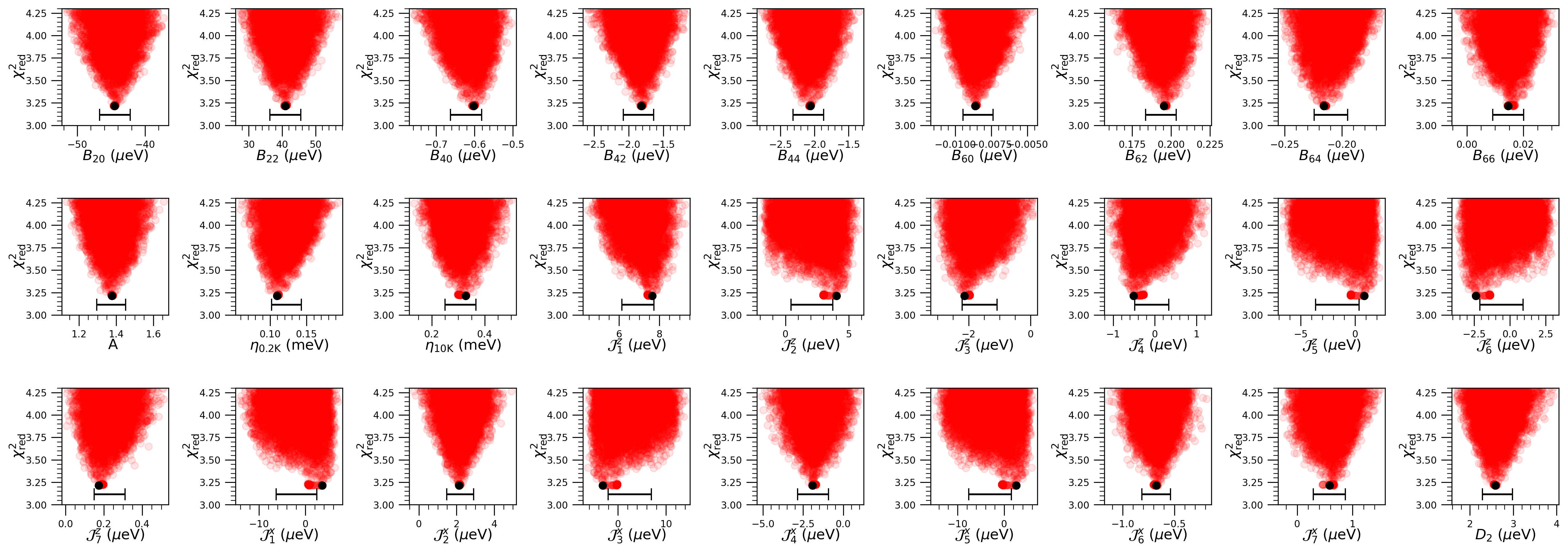}
    \caption{ Reduced chi-squared $\chi^2_\mathrm{red}$ metric for the parameters used in the RPA model. The black point is the best fit from the refinement. The parameters were varied to within $\Delta \chi^2_\mathrm{red} = 1$ of the minimum, and the reported values are given by the mean and standard deviation as shown in the erorr bar.}
    \label{fig:SI_ChiSq_PCA} 
\end{figure*} 

\begin{figure}[ht!]
    \centering
    \includegraphics[width=0.95\linewidth]{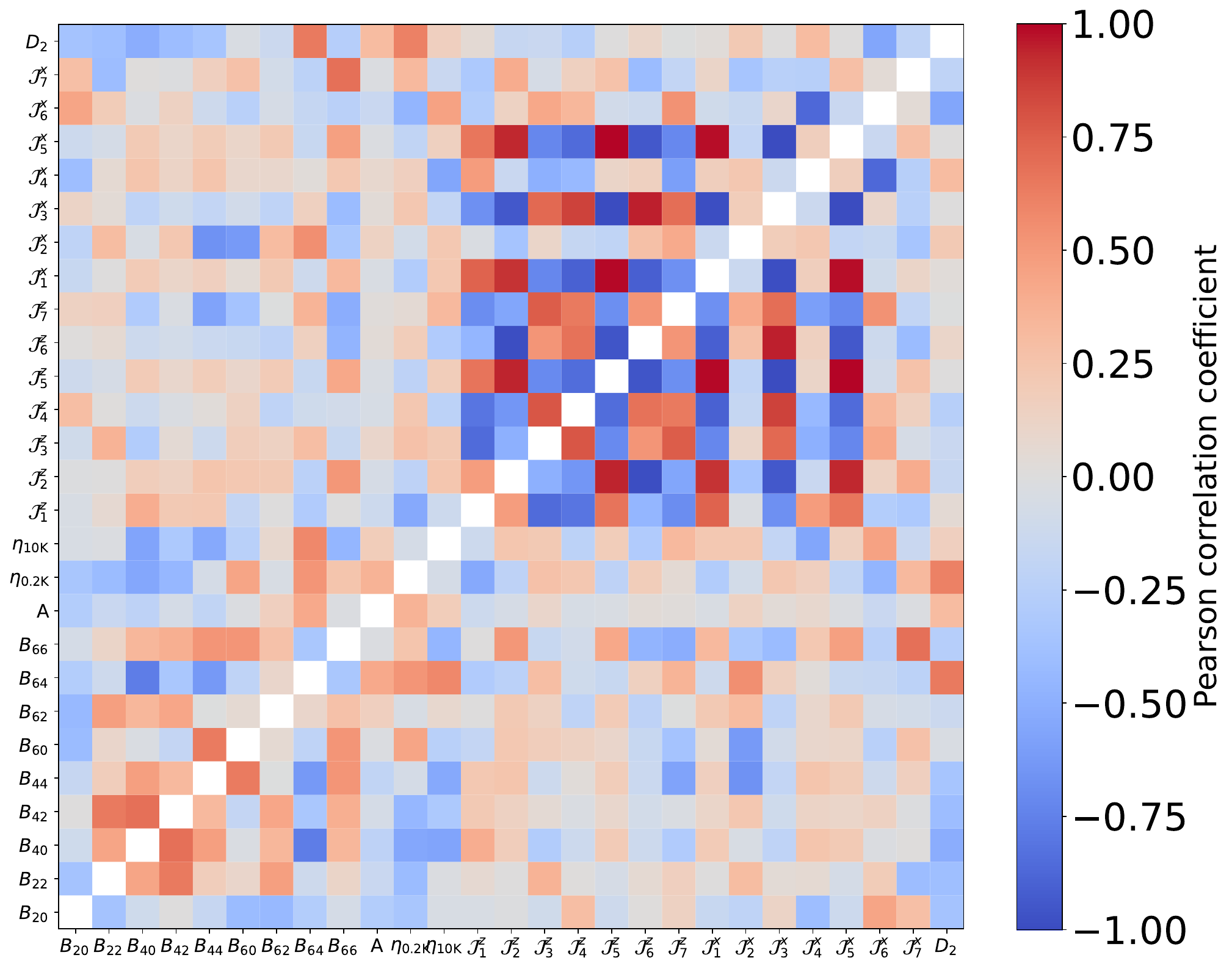}
    \caption{Correlations for the 27 parameters used in this model. It can be seen that most crystal field, amplitude, and width parameters are not highly correlated, whereas the exchange parameters have high correlations.}
    \label{fig:SI_correlations}
\end{figure}

Fig.~\ref{fig:SI_ChiSq_PCA} shows the reduced chi-squared versus the 27 independently refined parameters in the RPA model. These are the nine CEF parameters; the fourteen symmetric exchange parameters; the free antisymmetric exchange parameter $D_2$; the amplitude correction $A$; and the two width corrections $\eta$ at 10 K and 0.2 K. We also show the position of the mean and standard deviation used in the analysis. Note that many symmetric exchange parameters are highly correlated, as evidenced by computing the Pearson correlation coefficient for all parameters. These are shown in Fig.~\ref{fig:SI_correlations}. Sets like $\mathcal{J}_1^x, \mathcal{J}_3^x, \mathcal{J}_5^x$ are expected to be correlated due to their linear combination contributing to harmonics like $\cos 3\pi h$. Meanwhile, the crystal field and antisymmetric exchange parameters $D_2$ have separately low correlations. 

\subsection{Chirality and DM interaction}
In the case of spherical Weyl nodes with chirality $\chi_n$ and arranged in the Brillouin zone at position $\vb{Q}_n$, the contribution to the DM vector between spins is 
 \begin{equation*} 
    \vb{D}_{ij} \propto \vu{r}_{ij} \sum_{nm} \frac{\chi_n + \chi_m}{2} e^{i(\vb{Q}_n-\vb{Q}_m)\cdot \vb{r}_{ij}} 
 \end{equation*} 
This is strictly a bond-parallel interaction, and this model is strictly speaking independent of the symmetry of the bond. In this formalism, these are guaranteed by the symmetry of the \textit{Weyl nodes} in the Brillouin zone, considering their position and chirality. Note that for the bonds of interest in our model, in particular (i) the second-near neighbor bond along the $x$ or $y$ axes, or (ii) the fourth-near neighbor bond along the $(110)$ directions, the DM vectors will sum to zero identically. Hence, to a further approximation, it is of interest to consider the contribution of Weyl electrons for non-spherical Weyl nodes. 

Mirroring the derivation of Weyl-induced RKKY interactions between local moments, we split the induced spin interactions into "channels" defined by pairs $m, n$ of Weyl nodes in the 1st Brillouin zone. The low-energy Hamiltonian in the vicinity of a Weyl node at wavevector $\mathbf{Q}_{n}$ can be modeled as
\begin{equation} H_{n}(\mathbf{k}) = v \chi_{n} \vb*{\sigma} \cdot \vb{k}-\vb{u}_{n} \cdot \vb{k} - \mu \nonumber \end{equation} 
where $\chi_{n}$ is the chirality of the node $n, \mu$ is the chemical potential, and $\mathbf{k}$ is the electron's momentum relative to $\mathbf{Q}_{n}$. The tilt of the Weyl spectrum is parameterized by a velocity vector $\mathbf{u}_{n}$, and we assume $\left|\mathbf{u}_{n}\right|<v$ in order to focus on Type-I Weyl nodes. While various symmetry restrictions apply to $\chi_{n}, \mathbf{Q}_{n}$ and $\mathbf{u}_{n}$, here we consider electron scattering between general two Weyl nodes $n$, $m$ (assuming only that their node energy is the same, corresponding to $\mu=0$).

The effective spin Hamiltonian with induced RKKY interaction between local moments $\hat{\mathbf{n}}_{i}$ and $\hat{\mathbf{n}}_{j}$ at lattice sites $i, j$ (with coordinates $\mathbf{r}_{i}, \mathbf{r}_{j}$ ) has the form:
\begin{gather} H_{2}=\sum_{i j} J_{i j}^{a b} \hat{\mathbf{n}}_{i}^{a} \hat{\mathbf{n}}_{j}^{b} \nonumber \\ J_{ij}^{a b}=\frac{a^{6}}{2} \sum_{n m} \int \frac{d^{3} q}{(2 \pi)^{3}} e^{i\left(\mathbf{Q}_{m}-\mathbf{Q}_{n}+\mathbf{q}\right)\left(\mathbf{r}_{i}-\mathbf{r}_{j}\right)} K_{n m}^{a b}(\mathbf{q}) \end{gather}
Computing $K_{n m}^{a b}(\mathbf{q})$ in a single "channel" from the appropriate bubble Feynman diagram produces a form
\begin{multline}
    \delta K_{m n}^{a b}(\mathbf{q})=\frac{v\left(\chi_{m}+\chi_{n}\right) J_{K}^{2}}{(2 \pi)^{3}} i q \epsilon^{a b c} \\ \cdot \left[\hat{q}^{c} F^{\prime}\left(\frac{\Delta u_{m n}}{v}, \beta, q\right)+\hat{e}_{1}^{c} F^{\prime \prime}\left(\frac{\Delta u_{m n}}{v}, \beta, q\right)\right]
\end{multline}
in momentum space, where
\begin{gather}
    \Delta \mathbf{u}_{m n}=\mathbf{u}_{m}-\mathbf{u}_{n}=\Delta u_{m n}\left(\hat{\mathbf{q}} \cos \beta+\hat{\mathbf{e}}_{1} \sin \beta\right) \nonumber \\ \Delta u_{m n}=\left|\Delta \mathbf{u}_{m n}\right|
\end{gather} 

Only the terms associated with the DM interactions from equal-chirality node pairs are shown in $\delta K$. The functions $F^{\prime}, F^{\prime \prime}$ are complicated and their evaluation is useful for deducing how exactly the DM interactions eventually depend on the distance between two spins in real space (will not be considered here). The key new ingredient, which did not exist in the case of perfectly symmetric Weyl nodes, is the term with $\hat{e}_{1}^{c}$. The vector $\vu{r}_{1}$ is orthogonal to $\mathbf{q}$, but lives in the plane made by $\mathbf{q}$ and $\Delta \mathbf{u}_{m n}$. Here $\beta$ is the angle between $\mathbf{q}$ and $\Delta \mathbf{u}_{m n}$.

When $\mathbf{q}$ is integrated out to find $J_{i j}^{a b}$ in real space, the induced DM vector has a component parallel to the bond between the interacting spins, but also acquires a component perpendicular to the bond direction. The DM vector from a single $m, n$ channel lives in the plane made by the real-space bond direction $\vu{r}$ and the tilting difference $\Delta \mathbf{u}_{m n}$. Both components exhibit algebraically-attenuated and sign-changing behavior as a function of the distance between the two spins. If $\gamma$ is the angle between the vectors $\vu{r}$ and $\Delta \mathbf{u}_{m n}$, then the bond perpendicular component vanishes when $\gamma \in\{0, \pi\}$ or $\Delta u_{m n}=0$. Very roughly (ignoring the dependence on $\mu$ and approximating the dependence on $\gamma$ ), the bond-perpendicular component of the DM vector is smaller than the bond-parallel component in the same channel by a factor of $\sim\left(\Delta u_{m n} / v\right) \sin \gamma$.

\subsection{Additional Scattering Data} 

\begin{figure}
    \centering
    \includegraphics[width=0.75\linewidth]{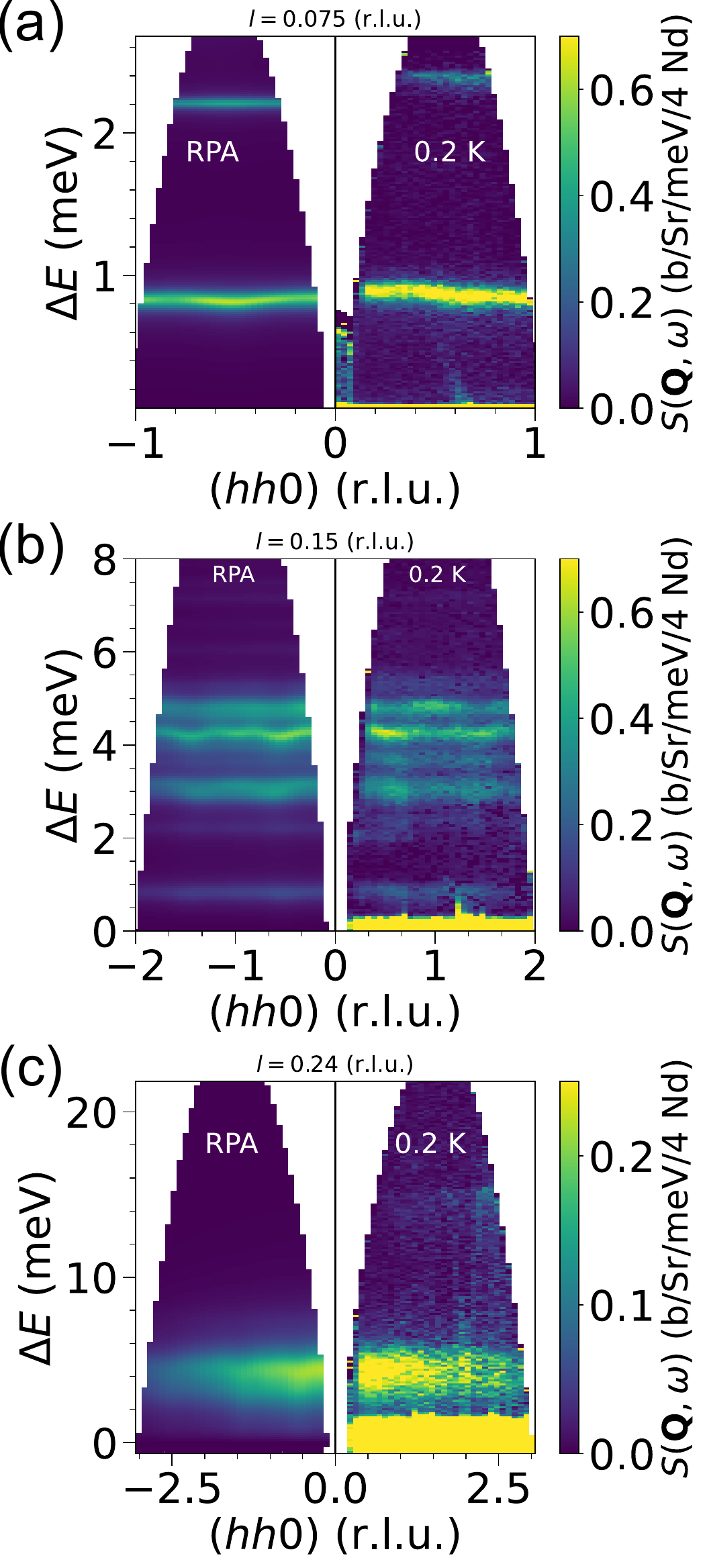}
    \caption{Commensurate state ($T=0.2$ K) simulation and data for the $(HH0)$ scattering plane at various incident energies for comparison. (a) $E_i = 3$ meV, (b) $E_i = 11$ meV, and (c) $E_i = 25$ meV.}
    \label{fig:SI_Different_Ei}
\end{figure}

Fig.~\ref{fig:SI_Different_Ei} shows the RPA simulation and dataset for $E_i=3$ meV, 11 meV (shown previously in Fig.~\ref{fig:RPA_Fitting}), and 25 meV. We note that only the 11 meV data were fit in the refinement, as this dataset had the best coverage of the excitons and sensitivity to the dispersion. The 3 meV data set has much higher resolution with comparatively less dispersion, but the consequence of it not being refined is that the $\omega_2 \sim 1.0$ meV mode is under-estimated in energy compared to that in the 11 meV dataset. Meanwhile, the 25 meV dataset has significantly broader resolution and therefore the excitons cannot be resolved. However, this dataset is useful to highlight that no extra modes appear above 10 meV in both the simulation and the data. The only high-energy features come from phonons at large $Q$ which have much stronger momentum dependence. 

\begin{figure}
    \centering
    \includegraphics[width=0.75\linewidth]{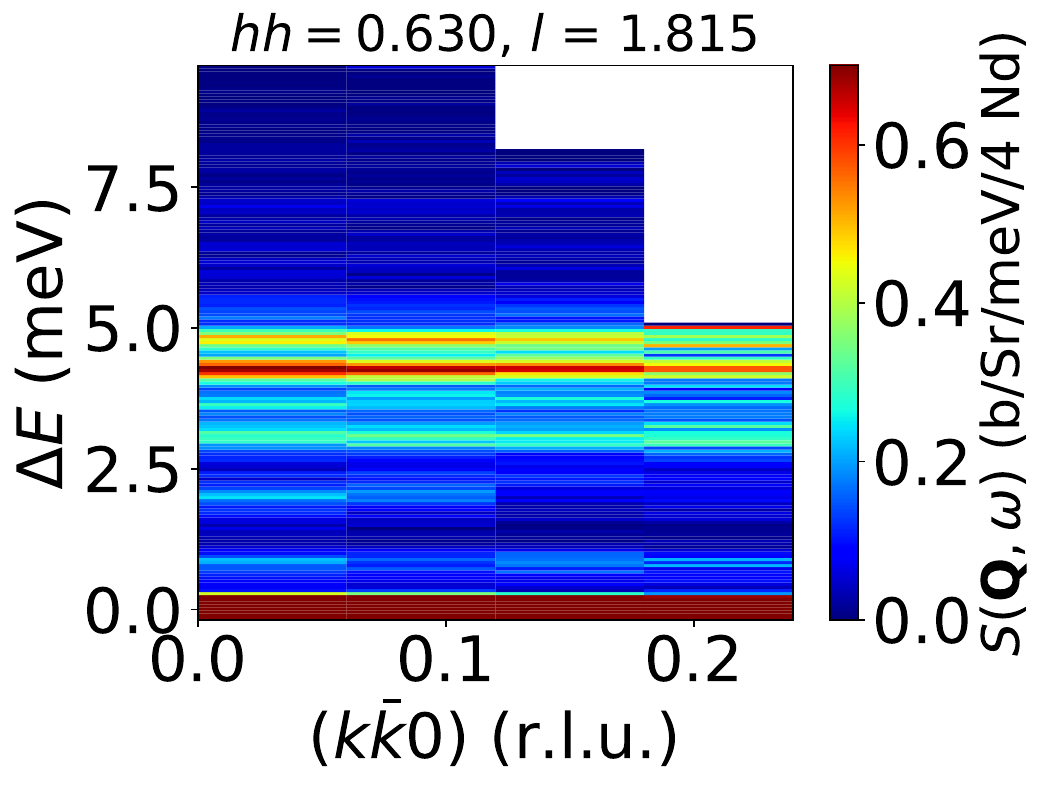}
    \caption{ Typical range of the scattering data from CNCS for $E_i=11$ meV, given the out-of-plane coverage, with a binning size $\Delta k = \Delta h$. The low statistics and small bandwidth makes fitting $K\neq 0$ prohibitive.}
    \label{fig:SI_KK0_range}
\end{figure}

Fig.~\ref{fig:SI_KK0_range} shows an example range of the $(k\bar{k}0)$ direction for selected values of $(hh0)$ and $(00l)$. The small out-of-plane coverage and small lattice constant $a$ means that the range of $(k\bar{k}0)$ is drastically smaller than that of $(hh0)$, making fits to nonzero $k\bar{k}$ difficult to refine. Therefore, the data we fit were integrated about $-0.1 < k\bar{k} < 0.1$ given the negligible dispersion in this direction. 

\begin{figure}
    \centering
    \includegraphics[width=\linewidth]{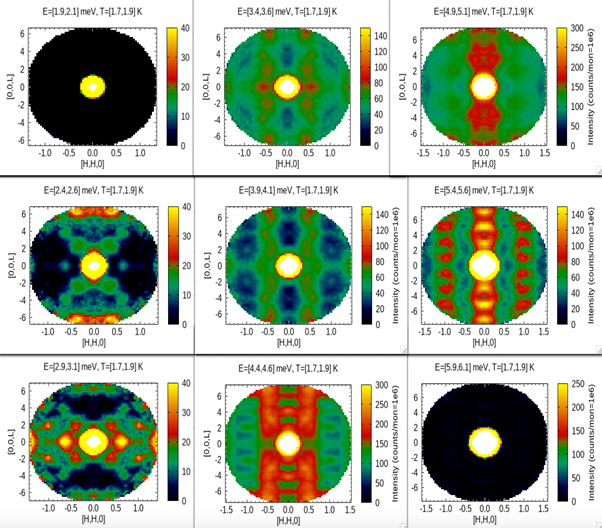}
    \caption{ MACS data in the $(hhl)$ scattering plane (same as CNCS) for various values of $\Delta E$. Note the integrated energy binning range of $\pm 0.1$ meV, compared with the finer energy binning $\delta E = 0.055$ meV enabled by the CNCS experiment. However, the salient features, present in both datasets, are highlighted here with improved momentum resolution.} 
    \label{fig:SI_MACS_data}
\end{figure} 

\bibliographystyle{apsrev4-1}
\bibliography{bibliography}

\end{document}